\documentclass[10pt, a4paper]{article}

\usepackage[intlimits]{amsmath}
\RequirePackage{snapshot}
\usepackage{graphicx}
\usepackage{pgfplots}
\usepackage[a4paper,margin=4cm]{geometry}
\usepackage[square,numbers]{natbib}
\usepackage{color}
\usepackage{esint}
\usepackage[mathscr]{euscript}
\usepackage{hyperref}
\usepackage{bbm}
\usepackage{lineno}
\usepackage{fancyhdr}
\usepackage[explicit]{titlesec}
\usepackage[labelfont=bf, font=normal]{caption}
\usepackage{tikz}

\usetikzlibrary{arrows,cd}

\usepackage[T1]{fontenc}

\newcommand{\rd}{\,\mathrm{d}}

\newcommand{\od}[2]{\dfrac{\mathrm{d} #1}{\mathrm{d} #2}}

\newcommand{\pd}[2]{\dfrac{\partial #1}{\partial #2}}
\newcommand{\vc}[1]{\boldsymbol{#1}}
\newcommand{\abs}[1]{\lvert{#1}\rvert}

\newcommand{\myheader}[0]{Stochastic models of ventilation driven by opposing wind and buoyancy}

\renewcommand{\subsectionmark}[1]{}

\titleformat{\subsubsection}[runin]{}{\thesubsubsection\quad\caps{#1}.}{1em}{}
\titleformat{\subsection}{\filcenter}{\thesubsection\quad\emph{#1}\indent}{1em}{}
\titleformat{\section}{\filcenter\bf}{\thesection\quad#1}{1em}{}
\titleformat{name=\section,numberless}[block]{\filcenter\bf}{}{1em}{#1}

\makeatletter
\newcommand{\specialnumber}[1]{%
  \def\tagform@##1{\maketag@@@{(\ignorespaces##1\unskip\@@italiccorr\emph{#1})}}%
}
\newcommand{\specialeqref}[2]{\begingroup
  \def\tagform@##1{\maketag@@@{(\ignorespaces##1\unskip\@@italiccorr\emph{#2})}}%
  \eqref{#1}\endgroup}
\makeatother

\begin{document}

\begin{center}
\vspace{-1cm}
\noindent{\LARGE\bf Stochastic models of ventilation driven\\ by opposing wind and buoyancy\par}\bigskip
{{\sc Veronica Andrian$^{1}$ and John Craske$^{1}$}\\
{\small $^{1}$Department of Civil and Environmental Engineering,\\ Imperial College London, London SW7 2AZ, UK}\\
   (Updated \today)}\\
\end{center}

\thispagestyle{empty}

\section*{Abstract}
\begin{quotation}
  \noindent Stochastic versions of a classical model for natural
  ventilation are proposed and investigated to demonstrate the
  effect of random fluctuations on stability and
  predictability. In a stochastic context, the well-known
  deterministic result that ventilation driven by the competing
  effects of buoyancy and wind admits multiple steady states can
  be misleading. With fluctuations in the buoyancy exchanged with
  an external environment modelled as a Wiener process, such
  systems tend to reside in the vicinity of global minima of their
  potential, rather than states associated with metastable
  equilibria. For a heated space with a leeward low-level and
  windward high-level opening, sustained buoyancy-driven flow
  opposing the wind direction is unlikely for wind strengths
  exceeding a statistically critical value, which is slightly larger
  than the critical value of the wind strength at which
  bifurcation in the deterministic system occurs. When
  fluctuations in the applied wind strength are modelled as an
  Ornstein-Uhlenbeck process, the topology of the system's
  potential is effectively modified due to the nonlinear role that
  wind strength has in the equation for buoyancy
  conservation. Consequently, large fluctuations in the wind of sufficient
  duration rule out the possibility of sustained ventilation
  opposing the wind direction at large base wind strengths.
\end{quotation}

\tableofcontents
  
\section{Introduction}\label{sec1}

By definition, buildings that are naturally ventilated are
subjected to forces from buoyancy and wind that fluctuate
unpredictably in time \cite{LinPafm1999a, HunGjfm2005a,
  HunGben1999a}. Examples include the heat and motion from a
variable number of occupants, transient heat sources from
equipment and solar gain, doors and windows that are opened or
closed intermittently, and the turbulence associated with external
wind loads
\cite{OUF2019257,HONG2016694,DELZENDEH20171061,SUN2017383,laaroussi2019occupant,
  boettcher2003statistics, CocOblm2006a, EftGblm2017a}. Indeed,
the role played by uncertainty has recently been highlighted more
broadly in the context of the succession events that lead to
COVID-19 transmission indoors \cite{BurHprs2020a,
  BhaRjfm2020a}. From the perspective of the deterministic
macroscopic models that are typically used to describe ventilation
mathematically, the examples listed above are random processes and
must therefore be treated stochastically.

In spite of the uncertainty in the environments to which they must be
applied, most existing ventilation models are deterministic and
implicitly average over fluctuations that would otherwise occur on
relatively short time scales. Such models include emptying filling
box models, which have proven to be useful for design and
successfully capture the leading-order physics behind ventilation
\cite{LinPafm1999a}. Their elegance
and simplicity belie the complex effects of turbulence for which
their parameters must necessarily account. Examples of such
parameters include the entrainment coefficient that is used to
model plumes in filling-box models \cite{BatGqms1954a,
  LinPjfm1990a, BaiWjfm1969a} and discharge coefficients that are
used to account for dissipation across openings
\cite{KarPijv2004a}.

In contrast to the deterministic models described in the previous
paragraph, there is growing recognition that probabilistic methods
are useful, if not essential, in analysing and predicting the
performance of buildings \cite{VesRbae2023a, LepJeab2022a}. For
modelling occupancy \cite{WolSape2019a}, passive tracers
\cite{MacMmea2018a} and wind speeds \cite{EftGblm2017a, MaJene2018a}, such
methods are already used. However, the incorporation of stochastic
processes in models for the fluid mechanics of natural ventilation is limited,
particularly in comparison with fields such as meteorology and
climate modelling \cite{PalTprs2008a}. 

\begin{figure}[t]
  \begin{center}
    \begin{tikzpicture}[scale=0.8]
      \node[align =left] at (2.5,3.5) {Forward flow};
      \fill[red,opacity = 0.2] (0,0)--(5,0)-- (5,3)--(0,3)--(0,0);
      \draw[] (0,0)-- (5,0) -- (5,2.7);
      \draw[] (0,0.3)-- (0,3) -- (5,3);
      \draw[red,->, very thick] (4.45,2.74) arc (115:85:2);
      \draw[red,->, very thick] (-0.35,0.1) arc (115:85:2);
      \node[align =left] at (2.5,1.5) {$b=b_3>0$};
      \draw[->] (0.1,-0.5) -- (0.05, -0.08); 
      \draw[->] (0.9,-0.5) -- (0.9, -0.08); 
      \draw[->] (1.7,-0.5) -- (1.7, -0.08); 
      \draw[->] (2.5,-0.5) -- (2.5, -0.08); 
      \draw[->] (3.3,-0.5) -- (3.3, -0.08); 
      \draw[->] (4.1,-0.5) -- (4.1, -0.08); 
      \draw[->] (4.9,-0.5) -- (4.9, -0.08); 
      \node[align =left] at (2.5,-0.78) {1};
      \draw[->] (6.2, 0) --(5.8,0);
      \draw[->] (6.2, 0.6) --(5.8,0.6);
      \draw[->] (6.2, 1.2) --(5.8,1.2);
      \draw[->] (6.2, 1.8) --(5.8,1.8);
      \draw[->] (6.2, 2.4) --(5.8,2.4);
      \draw[->] (6.2, 2.95) --(5.8,2.95);
      \draw[] (6.2,0) --(6.2,2.95);
      \node[align =left] at (6.6,1.5) {$W$};
    \end{tikzpicture}%
    \quad%
    \begin{tikzpicture}[scale=0.8]
      \node[align =left] at (2.5,3.5) {Reverse flow};
      \fill[red,opacity = 0.1] (0,0)--(5,0)-- (5,3)--(0,3)--(0,0);
      \draw[] (0,0)-- (5,0) -- (5,2.7);
      \draw[] (0,0.3)-- (0,3) -- (5,3);
      \draw[red,<-, very thick] (4.45,2.74) arc (115:85:2);
      \draw[red,<-, very thick] (-0.4,0.06) arc (115:85:2);
      \node[align =left] at (2.5,1.5) {$b=b_1<b_3$};
      \draw[->] (0.1,-0.5) -- (0.05, -0.08); 
      \draw[->] (0.9,-0.5) -- (0.9, -0.08); 
      \draw[->] (1.7,-0.5) -- (1.7, -0.08); 
      \draw[->] (2.5,-0.5) -- (2.5, -0.08); 
      \draw[->] (3.3,-0.5) -- (3.3, -0.08); 
      \draw[->] (4.1,-0.5) -- (4.1, -0.08); 
      \draw[->] (4.9,-0.5) -- (4.9, -0.08); 
      \node[align =left] at (2.5,-0.78) {1};
      \draw[->] (6.2, 0) --(5.8,0);
      \draw[->] (6.2, 0.6) --(5.8,0.6);
      \draw[->] (6.2, 1.2) --(5.8,1.2);
      \draw[->] (6.2, 1.8) --(5.8,1.8);
      \draw[->] (6.2, 2.4) --(5.8,2.4);
      \draw[->] (6.2, 2.95) --(5.8,2.95);
      \draw[] (6.2,0) --(6.2,2.95);
      \node[align =left] at (6.6,1.5) {$W$};
    \end{tikzpicture}
  \end{center}
  \caption{A space with lower and upper level openings, subjected
    to a nondimensionalised distributed heat source which, for
    deterministic models, is assumed to produce a well-mixed
    interior of uniform buoyancy (nondimensionalised as $b$ in
    \S\ref{sec:det}).  The buoyancy-driven ventilation is opposed
    by a pressure difference across each opening due to an
    opposing wind of (nondimensionalised) strength $W$. The
    left-hand figure illustrates `forward flow' and the right-hand
    figure illustrates `reverse flow' (the latter being possible
    when $W\geq 1$).}
    \label{fig:mixvent_sketch}
\end{figure}
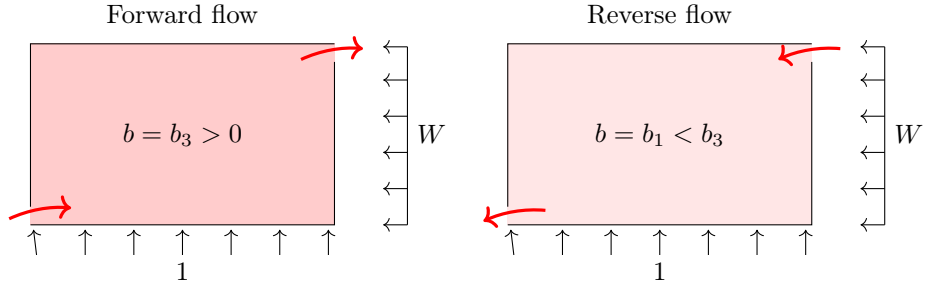

Some stochastic effects in natural ventilation have previously
been proposed for an example involving buoyancy and wind
\cite{FonAeab2013a}, which motivates the present study. The
example consists of a single room with low and high level openings
subjected to internal heating that produces a uniform temperature
and a wind that opposes flow out of the high-level opening (see
figure \ref{fig:mixvent_sketch}). By feeding samples from
probability distributions to account for occupancy, heating from
equipment and wind loads into equations describing a steady state,
previous work observed the effect of uncertainty in parameters on
the structure of the system's solutions and their sensitivity
\cite{FonAeab2013a}. However, the problem was not formulated as a
stochastic differential equation and analytical information that
can be obtained from the associated Fokker-Planck equation was not
utilised. More recently, stochastic fluctuations have been applied
to the wind opposing the displacement ventilation driven by a
point source of buoyancy \cite{VesRbae2023a}. Due to the nonlinear
dependence of the ventilation on the opposing wind strength, the
incorporation of fluctuations in the wind modifies the interface
height associated with the resulting two-layer stratification.

The extent to which existing bulk models of ventilation implicitly
account for random fluctuations is dependent on the time scales
under consideration and therefore moot; while the effects of small
scale turbulence and dissipation are usually incorporated
implicitly into discharge coefficients \cite{KarPijv2004a}, slowly
increasing wind loads or changes in occupancy would typically
appear explicitly as a time-dependent forcing
\cite{coomaraswamy2011time}. Being deterministic, they describe a
single trajectory or realisation, rather than a complete
description of the probabilities associated with a given
observable, which would correspond to infinitely more information
about the process. Indeed, deriving the probability densities
associated with differential equations is difficult in general,
due to the `curse of dimensionality' \cite{BelRnas1959a}. Each
degree of freedom of the system corresponds to an argument of the
system's probability density, whose evolution is therefore
governed by a partial (rather than ordinary) differential equation
\cite{KamNboo2011a}. Bulk models for ventilation, however, are
particularly amenable to a probabilistic interpretation because they
typically involve a small number of dimensions.

A further motivation for adopting a probabilistic perspective
comes from the fact that ventilation driven by the combined
effects of buoyancy and wind can produce multiple equilibria
(i.e. statistically steady states)
\cite{li2001natural,CheTbae2005a,HunGjfm2005a}. In contrast to
systems with a single equilibrium, the eventual fate of systems
possessing multiple equilibria depends on their initial
state. More generally, in cases where a parameter can change, it
is possible for a system to transition between equilibria and
exhibit hysteresis \cite{StrSboo1996a, HakHboo2013a}. In such
cases it is possible to identify the smallest perturbation in the
wind that is required to produce the transition for a single
well-mixed space \cite{YuaJbae2008a, LisBbae2009a} and for the
stratified environment created by a point source of buoyancy
\cite{CraJjfm2019b}. Systems subjected to random forces therefore
have the propensity to flip between metastable equilibria
\cite{GraTjsm2017a, MeuCjsp1988a, KueCphd2011a}, as demonstrated
by the classical van der Pol oscillator subjected to white noise
\cite{BurKprs2003a, SchKnon1996a} and `tipping points' in climate
models \cite{MenAprs2021a}.

In a stochastic setting, the pictures of such bifurcations becomes
blurred \cite{MeuCjsp1988a}; the `skeleton' provided by
bifurcation diagrams is obscured by the `flesh' of probability
density functions that quantify the likelihood of finding the
system in a particular state and affect the parameter values at which
`bifurcation' (in a sense that will be defined precisely in
\S\ref{sec:methods}) is likely to occur \cite{NamNamc1990a}. Among the
possible solutions identified in the previous paragraph are some
that are more probable than others, which means that designers do
not have to choose indiscriminately between possible
outcomes. Indeed, with density functions replacing deterministic
values, the `flesh' of probability provides a rich and practical
picture of the system's behaviour, which allows one to relate the
amplification of input noise with the underlying energetics of the
system and to quantify the time one can expect to wait for a
system to transition between states \cite{KraHpha1940a,
  VirLnlm1992a}.

As a prototypical example for establishing a probabilistic picture
of natural ventilation driven by the competing effects of buoyancy
and wind, this study follows previous work\cite{FonAeab2013a,
  GlaCjfm2001a} and considers the single room with low and high
level openings described above and illustrated in figure
\ref{fig:mixvent_sketch}. Previous qualitative observations are
made precise with analytical results and a more rigorous
formulation of the problem as a stochastic differential equation
and associated Fokker-Planck equation. The broader aim of the work
is therefore to demonstrate that useful information and insight
can be obtained by incorporating probabilistic effects into
existing ventilation models using accessible and well-established
methods of stochastic analysis. At the same time, various semantic
challenges of developing a rigorous foundation for stochastic
models of ventilation are highlighted.

The governing equations for the deterministic model are presented
in \S\ref{sec:det} and supplemented with a stochastic buoyancy
flux in \S\ref{sec:stochastic}, which is extended to a `Langevin'
formulation in \S\ref{sec:orn} by representing the wind as an
Ornstein-Uhlenbeck process. Analytical results derived from the
associated Fokker Planck equations are discussed in
\S\ref{sec:solutions} (and in \S\ref{sec:orn} for the extension),
and include predictions of the stationary probability
density. Numerical integration of the stochastic equations to
obtain realisations of the system and their comparison with the
analytical results are discussed in
\textsection\ref{sec:simulations}. In \S\ref{sec:app} the results
are interpreted in the context of real applications and
conclusions are drawn in \S\ref{sec:conclusions}.

\section{Governing Equations}\label{sec:methods}
\subsection{Deterministic Formulation}
\label{sec:det}

Consider the confined space illustrated in figure
\ref{fig:mixvent_sketch} with an upper and a lower level
opening. The interior is subjected to a distributed heat source at
floor level of strength $\hat{F}$. In a deterministic setting, the
temperature (and, therefore, buoyancy $\hat{b}$) of the air inside
the space is assumed to be uniform \cite{GlaCjfm2001a,
  coomaraswamy2011time}. Using the Boussinesq approximation, the
system is governed by the following buoyancy conservation
equation:

\begin{equation}
  \hat{H}\hat{S}\od{\hat{b}}{\hat{t}} = \hat{F}-\hat{Q}\hat{b}=\hat{F}-\hat{A}\bigg\lvert\hat{b}\hat{H}-\frac{\Delta\hat{p}}{\hat{\rho}_{0}}\bigg\rvert^{1/2}\hat{b},
\label{eq:det_dim}
\end{equation}

\noindent where $\hat{H}$ is the height of the space, $\hat{S}$ is
the horizontal (cross-sectional) area, $\Delta\hat{p}$ is the
pressure difference induced by the wind, $\hat{A}$ is an effective
opening area \cite{LinPafm1999a}, $\hat{Q}$ is a volume flux
corresponding to the magnitude of the ventilation and
$\hat{\rho}_{0}$ is a reference density. The right-hand side of
\eqref{eq:det_dim} represents the difference between the buoyancy
that is added to the space from heating and the buoyancy that is
lost from the space due to ventilation, which itself depends on
the pressure difference arising from both buoyancy and wind, and
therefore makes the equation nonlinear.

When the wind-induced pressure difference is below the critical
value \cite{GlaCjfm2001a, HunGjfm2005a, li2001natural},

\begin{equation}
\frac{\Delta \hat{p}_{c}}{\hat{\rho}_{0}}:=\left( \frac{27\hat{H}^{2}\hat{F}^{2}}{4\hat{A}^{2}}\right)^{1/3},
\label{eq:pc}
\end{equation}

\noindent equation \eqref{eq:det_dim} admits a single stable
steady-state solution (for which $\rd \hat{b}/\rd \hat{t}=0$)
corresponding to forward flow. When the pressure difference is
greater than the critical value, \eqref{eq:det_dim} admits two
additional steady-state solutions: a stable and an unstable
solution corresponding to reverse flow in the direction of the
wind (see figure \ref{fig:mixvent_sketch}). In the parlance of
dynamical systems theory, the steady-state solutions are known as
`fixed points' of \eqref{eq:det_dim}.

By introducing the following dimensionless quantities,

\begin{equation}
b:=\left(\frac{4\hat{A}^{2}\hat{H}}{27\hat{F}^{2}}\right)^{1/3}\hat{b},\quad\quad
t:=\left(\frac{4\hat{A}^{2}\hat{F}}{27\hat{H}^{2}\hat{S}^{3}}\right)^{1/3}\hat{t},\quad\quad
W:=\frac{\Delta\hat{p}}{\Delta\hat{p}_{c}},
\specialnumber{a,b,c}
\label{eq:dims}
\end{equation}

\noindent equation \eqref{eq:det_dim} can be written as

\begin{equation}
  \od{b}{t} = a(b;W):=1-c\lvert b-W\rvert^{1/2}\,b,
\label{eq:det}
\end{equation}

\noindent where $c =\sqrt{27/4}$ and $W$, quantifying the relative
strength of the wind as a squared Froude number, is the only free
parameter.

The fixed points of \eqref{eq:det} correspond to either local
minima (if they are stable) or local maxima (if they are unstable)
of the potential function

\begin{equation}
  V_{B}(b; W):= -b +\frac{2c}{15}(3b+2W)(b-W)\abs{b-W}^{1/2},
  \label{eq:potential}
\end{equation}

\noindent which is defined, to within an arbitrary additive function of $W$, such that

\begin{equation}
\od{b}{t} = -\od{V_{B}}{b}.
\end{equation}

\begin{figure}[t]
    \centering
    \includegraphics[width =\textwidth]{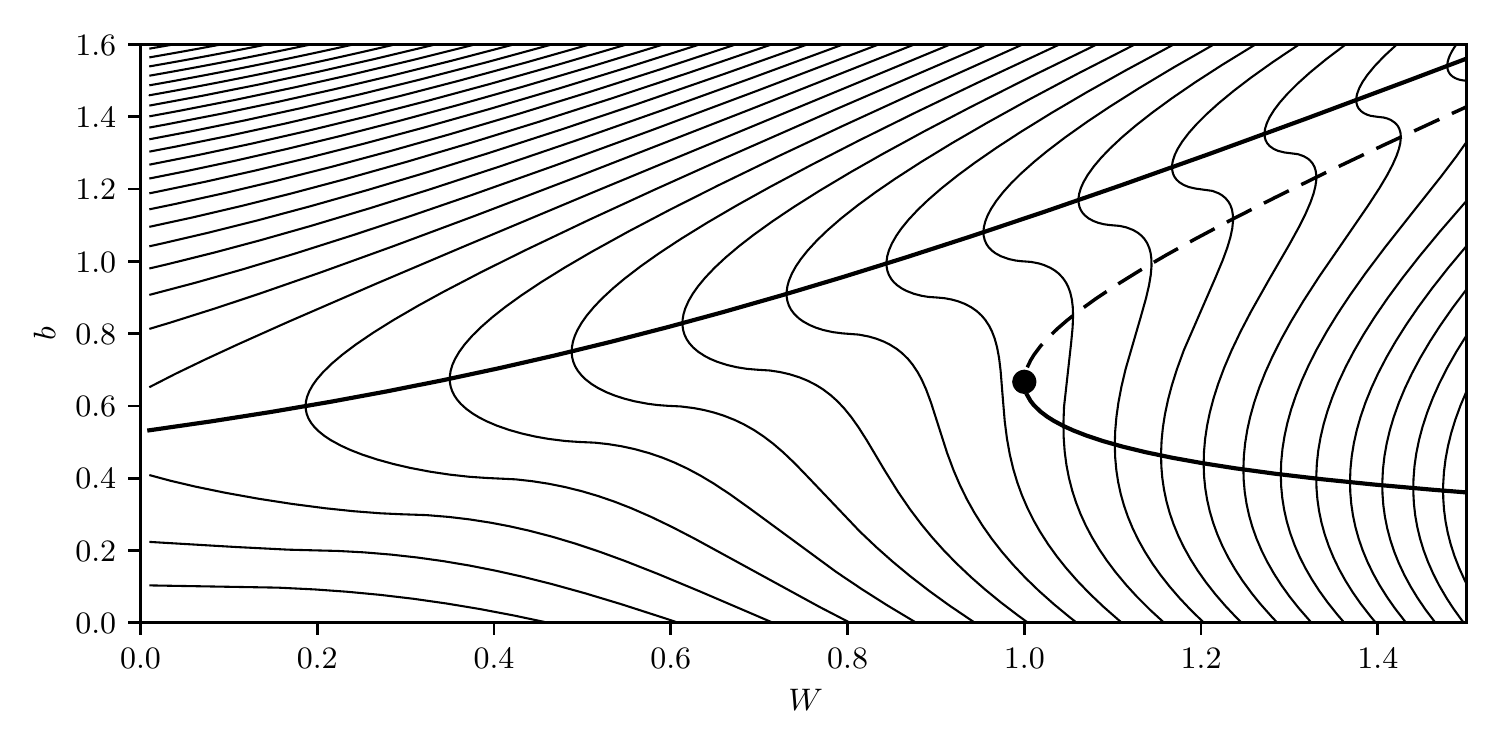}
    \caption{Bifurcation diagram corresponding to solutions of
      $a(b;W)=0$ and the underlying potential function $V_{B}(b;W)$
      for the deterministic system \eqref{eq:det}. The thick solid lines
      denote stable equilibria and the dashed line denotes
      unstable equilibria.}
    \label{fig:potential}
\end{figure}

Algebraically, the fixed points of
\eqref{eq:det} correspond to solutions to the cubic
equation $-\rd V_{B}/\rd b=a(b;W)=0$:
\begin{equation}
  b^{3}-Wb^{2}=\pm \frac{1}{c^{2}},
  \label{eq:fps}
\end{equation}

\noindent where the right-hand side is positive for forward flow
(one real solution, $b_{3}$) and, when $W\geq 1$, negative for
reverse flow (two real solutions, $b_{1}$ and $b_{2}$, where
$b_{1}<b_{2}<b_{3}$). Contours of $V_{B}(b;W)$ are shown beneath
the system's bifurcation diagram in figure
\ref{fig:potential}. The set of states that eventually end up at a
given fixed point is called the `basin of attraction' of that
fixed point, or, `potential well' in the context the energetics
implied by $V_{B}$.

\subsection{Stochastic (`Brownian') formulation}
\label{sec:stochastic}

There are many ways in which stochastic effects can be
incorporated in \eqref{eq:det}. However, in all cases it is
necessary to reappraise the meaning of the original deterministic
equation. Does \eqref{eq:det} represent an idealised system devoid
of disturbances or does it represent a system that has already
been averaged over sufficiently large time scales? Added to these
semantic difficulties of developing a stochastic model for
ventilation are the choices that must be made concerning the
physical origin of the fluctuations. While large scale
fluctuations in the wind would be expected to change the bulk
pressure difference acting on the building, smaller-scale
structures could produce bidirectional flows through the openings
that enhance buoyancy transfer without necessarily changing the
bulk ventilation. In this regard it should also be noted that, in
reality, the interior space is not of uniform buoyancy and that
temporal and spatial variations of buoyancy, and/or the heat
applied to the space, might also be treated as random variables.

If random noise is to be added to \eqref{eq:det} to create a
stochastic differential equation, which would provide trajectories
for individual realisations of the system, choices concerning the
rule used for time integration of the random fluctuations present
further technical and interpretative challenges
\cite{KamNjsp1981a}. Such `choices' are discussed in
\S\ref{sec:conclusions} and disappear altogether with sufficient
information about the physical properties of the noise and/or a
master equation describing the evolution of the system's
probability density.

In view of the challenges outlined above and the current lack of
fundamental justification for adopting any particular mathematical
form of the fluctuations that could affect \eqref{eq:det}, the
approach adopted here will be as simple and transparent as
possible. To account for the effects of rapid, uncertain and
external fluctuations applied to the system, a stochastic buoyancy
flux will be added to \eqref{eq:det}:

\begin{equation}
    \rd B_{t} = a(B_{t};W)\rd t + \sigma(B_{t}; W)\circ \rd \xi_{t},
\label{eq:sde}
\end{equation} 

\noindent where $B_{t}$ is the random variable associated with
buoyancy, $a(B_{t};W)$ is defined in \eqref{eq:det} and $\xi_{t}$
describes a Wiener process characterised by normally-distributed
jumps that are uncorrelated in time. More precisely, $\xi_{t}$ has
independent increments that are normally distributed, such that
$\xi_{t}-\xi_{s}\sim N(0,t-s)$ for $0\leq s<t$, and describes
continuous paths, with $\xi_{0}=0$ \cite{ResSboo1992a}. The
standard deviation of the fluctuations $\sigma(B_{t}; W)\geq 0$
appearing in \eqref{eq:det} can be adapted in accordance with the
physical origin of the fluctuations. In this regard it is
important to note that $\sigma(B_{t}; W)$ can account for
fluctuations in both the buoyancy and the velocity and, therefore,
does not rely on the assumption of uniform internal buoyancy that
is used for the deterministic model described in \S\ref{sec:det}. However,
significant further work is required to establish the
connection between fluctuations of buoyancy at an opening and the
internal/external conditions to which the space is subjected.

The Wiener process in \eqref{eq:sde} is integrated using the
Stratonovich interpretation \cite{KamNjsp1981a, MooNnjp2014a},
which, informally, means that $\sigma$ is evaluated at the
mid-point of each time step (see \S\ref{sec:simulations} for
further details). The Stratonovich interpretation used in
\eqref{eq:sde} represents an external source of fluctuations in
the limit of its autocorrelation time tending to zero
\cite{KamNjsp1981a}. This, and other subtleties associated with
the incorporation of stochastic effects in \eqref{eq:det} are
discussed further in \S\ref{sec:conclusions}.

Physically, the properties of $\xi_{t}$ correspond to the
fluctuations belonging to a normal distribution and occurring on
time scales that are significantly smaller than the integral
timescales associated with the deterministic version of the model
\eqref{eq:det}. In principle, $W$ and $\sigma$ could depend on
time explicitly to account for relatively slow and deterministic
properties of the environment. Here, however, it will be assumed
that \eqref{eq:sde} models the system's evolution for durations
over which $W$ and $\sigma$ can be treated as being independent of time.
As demonstrated in \S\ref{sec:orn}, as an extension
of the more tractable model considered in this section, finite time
correlations of fluctuations in the wind can be incorporated by
modelling $W$ in \eqref{eq:det} as an Ornstein-Uhlenbeck process
with an autocorrelation that decays exponentially in time
\cite{DooJaom1942a, UhlGphr1930a, VesRbae2023a}.

Given \eqref{eq:sde}, the probability density $f_{B}$ describes the
likelihood of observing a given buoyancy $B_{t}$, such that the probability

\begin{equation}
  \mathbbm{P}\{0\leq B_{t} \leq b\}=\int\limits_{0}^{b}f_{B}(\beta;t)\rd\beta.
\label{eq:P}
\end{equation}

The probability density evolves according to the Fokker-Planck equation \cite{RisHboo1996a}:

\begin{equation}
    \pd{f_{B}}{t} =- \pd{J_{B}}{b},
\label{eq:fp}
\end{equation}

\noindent where

\begin{equation}
  J_{B}:=a(b;W)f_{B}-\frac{\sigma}{2}\pd{}{b}(\sigma f_{B}),
\label{eq:buoyflux}
\end{equation} 

\noindent is a probability flux that satisfies
$J_{B}(0,t)=\lim\limits_{b\rightarrow\infty}J_{B}(b,t)=0\ \forall t$, which ensures that
probability is conserved in the domain, in the sense that
$\mathbbm{P}\{0\leq B_{t}<\infty\}=1\ \forall t$ for suitable
choices of the initial density. Indeed, equation \eqref{eq:fp} is
a `mass conservation' equation for probability and therefore
accounts for advection along the trajectory defined by
\eqref{eq:det} (typically referred to as `drift'), in addition to
diffusion due to the stochastic forcing in \eqref{eq:sde}. The
second term in the probability flux \eqref{eq:buoyflux} is
diffusion that results from the Stratonovich integration of the
noise in \eqref{eq:sde}. Integration of \eqref{eq:fp} against $b$
and commuting $\partial_{b}$ and $\sigma(b;W)$ using the product
rule in \eqref{eq:buoyflux} shows that the Stratonovich diffusion
term modifies the evolution of the mean buoyancy.

The solution of \eqref{eq:fp} yields the entire probability
density function for $B_{t}$ and corresponds to Einstein's
original formulation of Brownian motion \cite{EinAaph1905a}, in
which a stochastic force is applied directly to the `configuration
space' of a particle. In \S\ref{sec:orn}, on the other hand,
fluctuations will be applied to an additional `velocity space' for the
wind, which is analogous to Langevin's subsequent and more general
formulation \cite{LanPmis1908a, GilDpre1996a}. A further, independent,
difference between the two approaches is that while Einstein
presented his model in terms of a Fokker-Planck equation
(cf. \eqref{eq:fp}), Langevin manipulated the stochastic
differential equations describing particle trajectories directly
\cite{LemDajp1997a}.

\section{Analytical results}
\label{sec:solutions}

Equation \eqref{eq:fp} is a one-dimensional advection diffusion
equation and has analytical solutions that can be expressed in
terms of a potential function that is scaled to account for the
state-dependent variance $\sigma(B_{t};W)^{2}$ of the
destabilising random fluctuations. To derive the solution it is
first convenient to derive a potential function in terms of a new
random variable $X_{t}$, in terms of which the variance $\sigma^{2}$
becomes independent of state and, therefore, constant.

\subsection{Coordinate transformation}

The monotonic transformation $g:b\mapsto x$ that produces a Fokker-Planck
equation with a constant diffusivity $\varepsilon^{2}/2$ in the new
frame of reference is \cite{RisHboo1996a, HakHboo2013a}:

\begin{equation}
  g(b):=\int_{0}\limits^{b}\frac{\varepsilon}{\sigma(\beta; W)}\rd\beta,
  \label{eq:x}
\end{equation}

\noindent and the probability density (corresponding to a pushforward measure)
associated with the new variable $x$ satisfies

\begin{equation}
f_{X}(x,t)=f_{B}(g^{-1}(x),t)\od{}{x}g^{-1}(x),
\label{eq:BtoX}
\end{equation}

\noindent or, in the other direction,

\begin{equation}
f_{B}(b,t)=f_{X}(g(b),t)\od{}{b}g(b).
\label{eq:XtoB}
\end{equation}

\noindent The transformation $x=g(b)$ is simply a mathematical
device that squeezes and stretches the coordinate system for
buoyancy such that the probability density for $X_{t}$ satisfies
a Fokker-Planck equation with constant diffusivity
$\varepsilon^{2}/2$,

\begin{equation}
    \pd{f_{X}}{t} =- \pd{J_{X}}{x},
\label{eq:fpX}
\end{equation}

\noindent where

\begin{equation}
  J_{X} =-\left(\od{V_{X}}{x}+\frac{\varepsilon^{2}}{2}\pd{}{x}\right)f_{X},
  \label{eq:JX}
\end{equation}

\noindent for which the potential function (noting that $b$ must
be obtained by inverting \eqref{eq:x}) is defined to within an
arbitrary additive function $V_{0}$ of $W$ according to

\begin{equation}
  V_{X}:=-\int\limits^{x}\frac{\varepsilon}{\sigma}\left[1-c\abs{g^{-1}(x)-W}^{1/2}g^{-1}(x))\right]\rd x+V_{0}(W).
  \label{eq:Ex_general}
\end{equation}

\noindent Equation \eqref{eq:fpX} adopts a canonical frame of
reference for obtaining analytical solutions and is the setting of
the classical stochastic problem of determining the escape rate
from a potential well \cite{KraHpha1940a}, which will be examined in
\S\ref{sec:kramer}. The formulation shows immediately that local
minima and maxima of $V_{X}$ correspond to zeros of the term in
square brackets in \eqref{eq:Ex_general} and therefore coincide
with the fixed points of the original system \eqref{eq:det},
which, incidentally, would not have been the case if the
stochastic term in \eqref{eq:sde} had been integrated according to
the It\^{o} interpretation.

\subsection{Quadratic diffusion}
\label{sec:quad}

As a specific example of the model presented in
\S\ref{sec:stochastic}, assume that $\xi_{t}$ in \eqref{eq:sde}
characterises an `externally imposed' fluctuation in the
ventilation, such that the standard deviation of the forcing
applied to \eqref{eq:sde} (representing the strength of the
resulting fluctuations in the buoyancy flux) can be expressed as

\begin{equation}
  \sigma(b; W) = \varepsilon(W)b.
  \label{eq:sigma}
\end{equation}

\noindent For windward and leeward openings perpendicular to the
wind direction, one would expect $\varepsilon$ to increase with
respect to $W$. However, for oblique orientations it is
conceivable that large fluctuations could occur at relatively
small values of $W$. Noting these possibilities, the results below
will be presented for constant $\varepsilon$ but can be adapted
immediately for particular relationships between $W$ and $\varepsilon$.

According to \eqref{eq:Ex_general}, the potential function
associated with \eqref{eq:sigma} can be expressed as 

\begin{equation}
  V_{X}(x):= e^{-x}+2\left(Q-cW^{1/2}h\left(\frac{Q}{cW^{1/2}}\right)\right)+V_{0}(W),
\label{eq:Ex}
\end{equation}

\noindent where $e^{x}=:g^{-1}(x)=b$, $Q:=c\abs{b-W}^{1/2}$ and $h$ is a function that depends
on whether the flow is forward (first case) or reverse (second case):

\begin{equation}
  h :=
  \begin{cases}
    \mathrm{arctan},\quad\quad &e^{x}=b \geq W,\\
    \mathrm{arctanh},\quad\quad &e^{x}=b < W.
  \end{cases}
\end{equation}

\begin{figure}[t]
    \centering
    \includegraphics{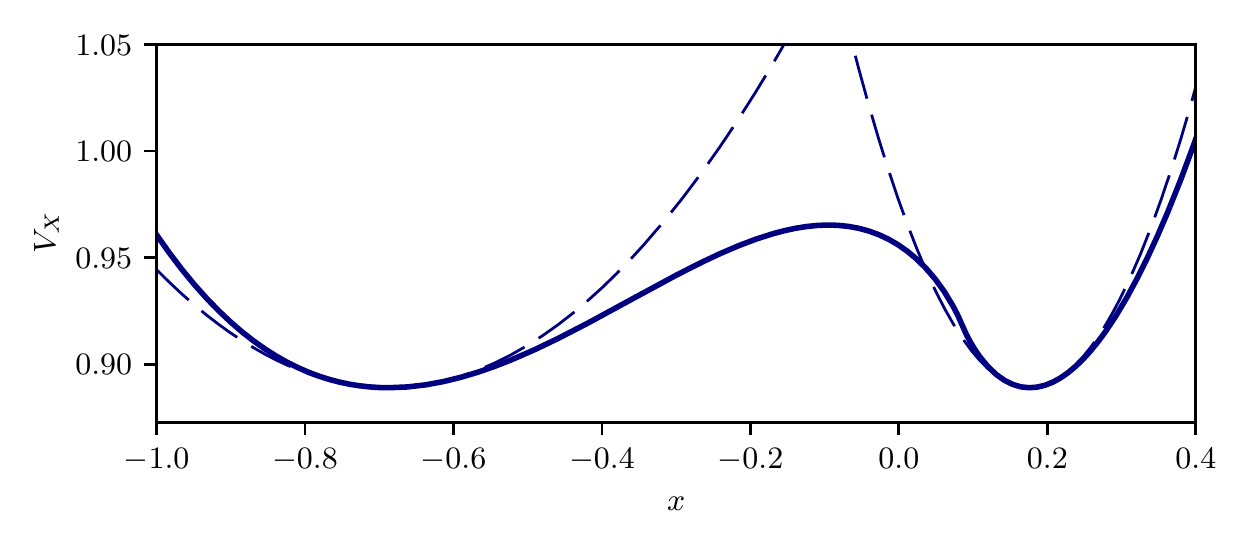}
    \caption{Parabolic approximations corresponding to the first
      two terms of \eqref{eq:expansion} (dashed lines) to the
      exact potential \eqref{eq:Ex} (solid line) at
      $W=W_{s}=1.0887$, which is discussed in \S\ref{sec:ws}, for $\sigma(b;W)=\varepsilon b$.}
    \label{fig:parabola}
\end{figure}

\noindent Although the constant $V_{0}(W)$ is arbitrary and therefore
does not affect the dynamics induced by $V_{X}$, selecting
$V_{0}(W)$ to ensure that

\begin{equation}
\min\limits_{x}V_{X}(x;W)=0, 
\end{equation}

\noindent is useful for the numerical conditioning of expressions
involving $V_{X}$ below (particularly \eqref{eq:fB}). For plotting
$V_{X}$, however, it will often be convenient to adopt
$V_{0}(W)\equiv 0$ (cf. \eqref{eq:potential} and figure
\ref{fig:potential}). The potential \eqref{eq:Ex} is illustrated
in figure \ref{fig:parabola} (with $V_{0}(W)\equiv 0$), along with
parabolic approximations in the vicinity of each local minima that
will be discussed in \S\ref{sec:stationary}.

The local minima of $V_{X}$ occur at the stable fixed points
$b_{1}$ and $b_{3}$ (denoted $x_{1}$ and $x_{3}$ in the
$x$-domain, respectively) of \eqref{eq:det}. Use of
\eqref{eq:sigma} implies that $V_{X}\rightarrow \infty$ as
$b\rightarrow 0$ ($x\rightarrow -\infty$), due to the first term
on the right-hand side of \eqref{eq:Ex} and
$V_{X}\rightarrow \infty$ as $b\rightarrow \infty$
($x\rightarrow \infty$), due to the remaining terms. Consequently,
as will be seen in the next section, the value and gradient of the
probability density in these limits tends to zero and therefore satisfies the
boundary conditions for the buoyancy flux discussed below \eqref{eq:buoyflux}.

\subsection{The stationary (invariant) density}
\label{sec:stationary}

In a statistically stationary state $\partial_{t}f_{X}\equiv 0$,
which, given the boundary conditions $J_{X}(0,t)=\lim\limits_{b\rightarrow\infty}J_{X}(b,t)=0\ \forall t$,
implies a `detailed balance' \cite{RisHboo1996a} in which the probability
flux \ref{eq:JX} vanishes everywhere:

\begin{equation}
  \pd{f_{X}}{x}=-\frac{2}{\varepsilon^{2}}\od{V_{X}}{x}f_{X}.
  \label{eq:stationary}
\end{equation}

\noindent If $f_{X}^{*}$ is the probability density (stationary
densities will be denoted with `$*$') satisfying
\eqref{eq:stationary}, then

\begin{equation}
  f_{X}^{*}(x;W)=N\,\mathrm{exp}\left(-\frac{2V_{X}(x;W)}{\varepsilon^{2}}\right),
  \label{eq:fB}
\end{equation}

\noindent for which $N$ is the choice of integration constant that
normalises $f^{*}_{X}$ (i.e. ensures that it is a probability density
whose integral with respect to $x\in (-\infty,\infty)$ is equal to
$1$) and $f_{B}^{*}$ is obtained from \eqref{eq:BtoX}. Equation
\eqref{eq:fB} is equivalent to the Boltzmann distribution in
statistical mechanics \cite{LanLboo1959a}. As can be seen from
\eqref{eq:fB}, since $\varepsilon$ is a constant, local minima and
maxima in $f_{X}^{*}$ correspond to local maxima and minima in
$V_{X}$, respectively, although the same cannot be said about the
relationship between $f_{B}^{*}$ and $V_{B}$ due to the pushforward
transformation rule \eqref{eq:BtoX}.

\begin{figure}[t]
    \centering
    \includegraphics[width =\textwidth]{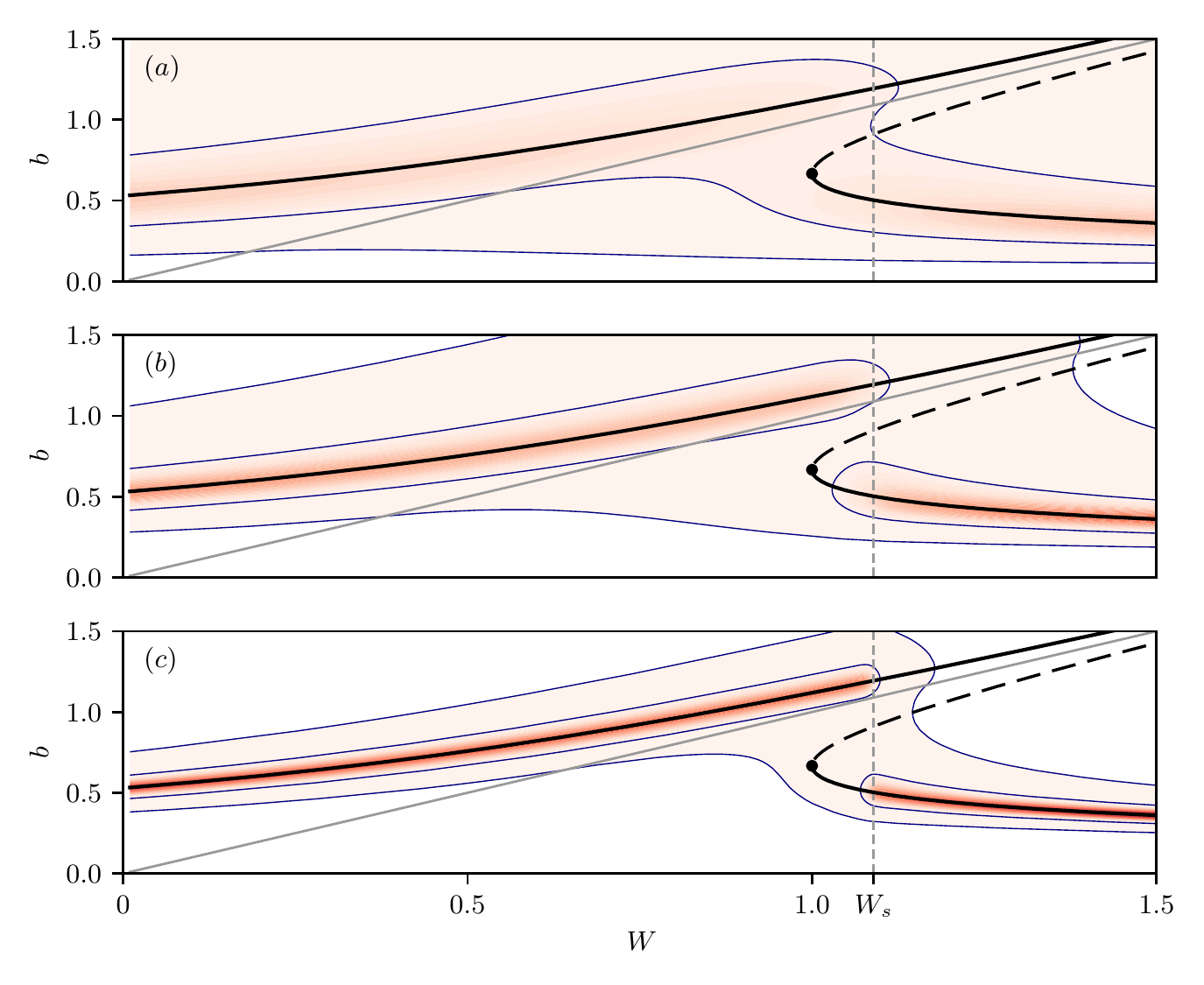}
    \caption{The stationary probability density $f_{B}^{*}$ (shades of red) as
      a function of the wind strength $W$ when
      $\sigma(b;W)=\varepsilon b$ for $\varepsilon=0.5\ (a)$,
      $\varepsilon=0.25\ (b)$ and $\varepsilon=0.125\ (c)$. The thin blue lines highlight contours corresponding to a stationary density of $10^{-8}$ and $5\times 10^{-1}$. As demonstrated in \S\ref{sec:ws}, when
      $\varepsilon\rightarrow 0$ the system resides in a state of
      forward flow for $W<W_{s}$ and reverse flow for
      $W_{s}<W$. The solid grey line corresponds to $b=W$ and
      therefore separates points corresponding to forward flow
      (above) and reverse flow (below).}
    \label{fig:stationary}
  \end{figure}

Figure \ref{fig:stationary} illustrates the dependence of the
stationary probability density \eqref{eq:fB} on the buoyancy $b$
and wind strength $W$ (in this regard, $f^{*}_{B}$ should be viewed as
a probability density conditional on a given value of $W$).  As
the strength of the noise decreases, the probability density
becomes concentrated in narrow bands around the deterministic
system's bifurcation diagram. At a particular value of the wind
strength, which will be denoted $W_{s}$ and defined precisely in
\S\ref{sec:ws}, a transition in the most likely state of the
system occurs.  When $W<W_{s}$ the system is predominantly found
in forward flow, whereas when $W>W_{s}$ the system is
predominantly found in reverse flow. Rather than ascribing equal
probability to the forward and reverse branches of the bifurcation
diagram, the stationary probability density accounts for the
tendency of the system to minimise the underlying
potential \eqref{eq:Ex}. In addition, a crucial feature of the
stationary probability density in the limit
$\varepsilon\rightarrow 0$ is that the point $W=W_{s}> 1$, at which
the system transitions from forward to reverse flow, does not
correspond to the point of bifurcation in the underlying
deterministic system.

\subsection{Local approximation}
  
In the vicinity of a local minimum at $x=x_{m},\ m\in \{1,3\}$, of the potential
$V_{X}$, it is useful to scale the excursions $\Delta x_{m}:=x-x_{m}$ by
the magnitude of the noise and the restoring force provided by the
potential well:

\begin{equation}
  \Delta y_{m} :=\sqrt{V_{X}''(x_{m})}\frac{\Delta x_{m}}{\varepsilon}.
\label{eq:dy}
\end{equation}

\noindent The definition of $\Delta y_{m}$ allows the potential function to be expressed as  

\begin{equation}
  V_{X}(x)\sim V_{X}(x_{m})+\frac{\varepsilon^{2}}{2}\Delta y_{m}^{2}+\frac{\varepsilon^{3}\Delta y_{m}^{3}}{3!}\frac{V_{X}'''(x_{m})}{V_{X}''(x_{m})^{3/2}}+\ldots,
  \label{eq:expansion}
\end{equation}

\noindent from which the term of order $\varepsilon$ is missing
because the first derivative $V_{X}'$ is zero at the local minimum
at $x_{m}$ by definition.

Under the assumption that

\begin{equation}
\od{^{n}V_{X}}{x^{n}} \left(\od{^{2}V_{X}}{x^{2}}\right)^{-\frac{n}{2}}=o\left(\varepsilon^{2-n}\right)\quad \mathrm{as\ } \varepsilon\rightarrow 0\ \ \mathrm{at\ \ } x=x_{m}\ \ \mathrm{for\ \ }n=3,4,\ldots,
\end{equation}

\noindent all but the first two terms in the expansion
\eqref{eq:expansion} can be discarded as being relatively small in
the limit $\varepsilon\rightarrow 0$, leading to the approximation
of the potential function as a set of parabolas. Locally, the
expansion therefore produces Gaussian probability densities with
respect to $x$, conditional on the chosen minimum $m$:

\begin{equation}
f^{*}_{X|M}(x;m)\sim \sqrt{\frac{V''(x_{m})}{\varepsilon^{2}\pi}}\mathrm{exp}\left(-\Delta y_{m}(x)^{2}\right).
\label{eq:fXM}
\end{equation}

\noindent The contribution that the local density \eqref{eq:fXM}
makes to $f_{B}^{*}$ can be calculated by applying the
transformation \eqref{eq:XtoB} and scaling the result by the
probability of observing $B_{t}^{*}\in \mathscr{B}_{m}$ for each
$m$, where $B_{t}^{*}$ refers to the statistically stationary
buoyancy:

\begin{equation}
  f_{B}^{*}(b)\sim \sum\limits_{m=1,3}\mathbbm{P}\{B^{*}_{t}\in\mathscr{B}_{m}\}f^{*}_{X|M}(g(b);m)\od{}{b}g(b),
  \label{eq:fBa}
\end{equation}

\noindent where $\mathscr{B}_{m}$ is the basin of attraction associated
with the stable fixed point at $b_{m}$ and
$\mathbbm{P}\{B^{*}_{t}\in\mathscr{B}_{m}\}$ is the corresponding
approximation to the probability of finding the statistically
stationary buoyancy in the basin of attraction associated with the
fixed point $m$ as $\varepsilon\rightarrow 0$:

\begin{equation}
  \mathbbm{P}\{B^{*}_{t}\in\mathscr{B}_{m}\}\sim N\sqrt{\frac{\varepsilon^{2}\pi}{V_{X}''(x_{m})}}\mathrm{exp}\left(-2\frac{V_{X}(x_{m})}{\varepsilon^{2}}\right).
\end{equation}

\noindent In this regard, the normalisation constant $N$ is equal
to the constant appearing in \eqref{eq:fB} as
$\varepsilon\rightarrow 0$ and should be chosen to ensure that
$\mathbbm{P}\{B^{*}_{t}\in\mathscr{B}_{1}\}+\mathbbm{P}\{B^{*}_{t}\in\mathscr{B}_{3}\}=1$. In
order to satisfy the zero probability flux boundary condition at
$b=0$ exactly, an `image' solution located at
$b=-b_{m}=-g^{-1}(x_{m})$ could be added to
\eqref{eq:fBa}. However, since $\varepsilon$ is required to be
small in order for \eqref{eq:fBa} to be valid, the error that
would otherwise be induced by \eqref{eq:fBa} at the boundary $b=0$
will be exponentially small.

\begin{figure}[t]
  \begin{center}
    \includegraphics[]{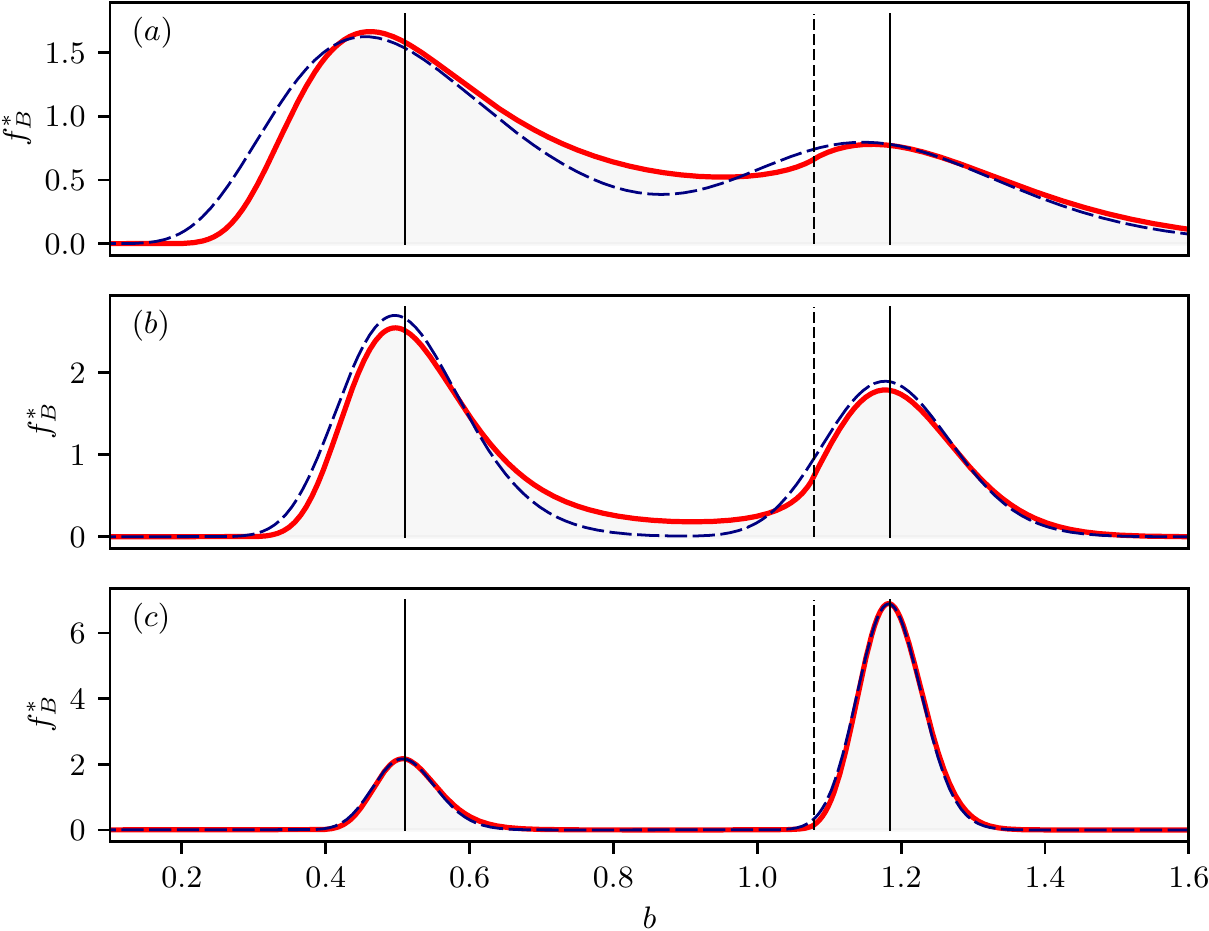}
  \end{center}
  \caption{The stationary probability density at $W=W_{s}-0.01$
    (which means that
    $\mathbbm{P}\{B_{t}^{*}\in\mathscr{B}_{1}\}\rightarrow 0$ as
    $\varepsilon\rightarrow 0$) when $\sigma(b;W)=\varepsilon b$
    for $\varepsilon=0.5\ (a)$, $\varepsilon=0.25\ (b)$ and
    $\varepsilon=0.125\ (c)$. The thick red line denotes the exact
    solution \eqref{eq:fB} and the blue dashed line denotes the
    Gaussian approximation that becomes increasingly accurate as
    $\varepsilon\rightarrow 0$. The vertical dashed lines
      separate reverse flow ($b<W$) from forward flow ($W<b$). The vertical black lines indicate
    the location of the stable fixed points for forward flow and
    reverse flow.}
    \label{fig:approximation}
\end{figure}

Figure \ref{fig:approximation} displays the stationary probability
density $f^{*}_{B}$ and the approximation \eqref{eq:fBa} for
different values of $\varepsilon$ at a wind strength just below
$W_{s}$. When $\varepsilon=0.5$, figure \ref{fig:approximation}
suggests that the typical excursions of the system (characterised
by the ratio of $\varepsilon$ and the restoring force provided by
the curvature of a potential well) are at least comparable to the
size of each basin of attraction. Indeed, evaluation of $V_{X}''$
implies that when $\varepsilon=0.5$,
$\varepsilon/\sqrt{V_{X}''(x_{1})}\approx 0.5$, which is close to
the distance between the local minimum of $V_{X}$ associated with
reverse flow and the local maximum of $V_{X}$ (see figure
\ref{fig:parabola}). Consequently, the likelihood of the system
being found in the vicinity of the local maximum of $V_{X}$ that
separates the two potential wells is not insignificant, which
suggests that the system flips regularly between forward and
reverse flow. In such situations \eqref{eq:fBa}, which essentially
assumes that the system is composed of isolated quadratic
potential wells, is not an accurate approximation. However, as
$\varepsilon$ is reduced to $0.125$, the probably density
associated with the unstable fixed point (the local maximum of
$V_{X}$) approaches zero and the approximation \eqref{eq:fBa}
becomes almost indistinguishable from the stationary probability
density \eqref{eq:fB}, which consists of two disjoint densities
centred on the potential wells.

\subsection{The (statistically) critical wind strength}
\label{sec:ws}

Figures \ref{fig:potential} and \ref{fig:stationary} suggest that
the asymmetry of the system's underlying potential biases the
probability density that describes its eventual state. In
particular, as $\varepsilon\rightarrow 0$, the exponent on the
right-hand side of \eqref{eq:fB} is dominated by the minimum value
of $V_{X}$ and converges weakly to a dirac measure at the point at
which the minimum occurs (assuming it is unique). At small values
of $W$, the deepest potential well corresponds to forward flow,
whereas, for large values of $W$, the deepest potential well
corresponds to reverse flow. Between these two regimes, for some
intermediate value $W_{s}>1$, the depths of the potential wells
are equal and the dirac measure, to which the distributions with
$\varepsilon\rightarrow 0$ converge, switches from forward flow to
reverse flow.

Stated differently, if one is prepared to wait long enough, the
state of the system subjected to asymptotically vanishing noise
will be found in the deepest potential well. It is therefore
important to know the value $W_{s}$ that renders the two potential
wells of equal depth, since such a value will correspond to the
transition in the expected state from forward flow to reverse flow.

The required definition of $W_{s}$ is:

\begin{equation}
  \lim\limits_{\varepsilon\rightarrow 0}\mathbbm{P}\{B_{t}^{*}\in\mathscr{B}_{1}(W)\}=
  \begin{cases}
    0,&\quad\quad W<W_{s},\\
    1,&\quad\quad W>W_{s},
  \end{cases}
\end{equation}

\noindent which states that the statistically stationary system
will almost always be in the basin of attraction for forward or
reverse flow as $\varepsilon\rightarrow 0$ when $W<W_{s}$ and
$W>W_{s}$, respectively. The `statistically' critical value of the
wind $W_{s}=1.0887$ makes the depth of the potential wells for
reverse flow and forward flow equal, and therefore satisfies

\begin{equation}
  V_{X}(x_{1}(W_{s}))=V_{X}(x_{3}(W_{s})).
\end{equation}

\noindent In spite of the fact that forward flow is a stable fixed
point of the underlying deterministic system, from the perspective
of the system's energetics, and in the context of practical
applications, one should expect reverse flow to eventually
dominate the system's response for sufficiently large opposing
wind strengths. Globally, the fixed point associated with forward
flow for $W>W_{s}$ corresponds to a point of metastable
equilibrium, which, since the stochastic forcing does not have
compact support, is not stable enough to prevent an eventual
transition to reverse flow. The next section quantifies how long
one would typically need to wait for this `eventual' transition
between forward flow and reverse flow to occur.

\begin{figure}[t]
    \centering
    \includegraphics[width = \textwidth]{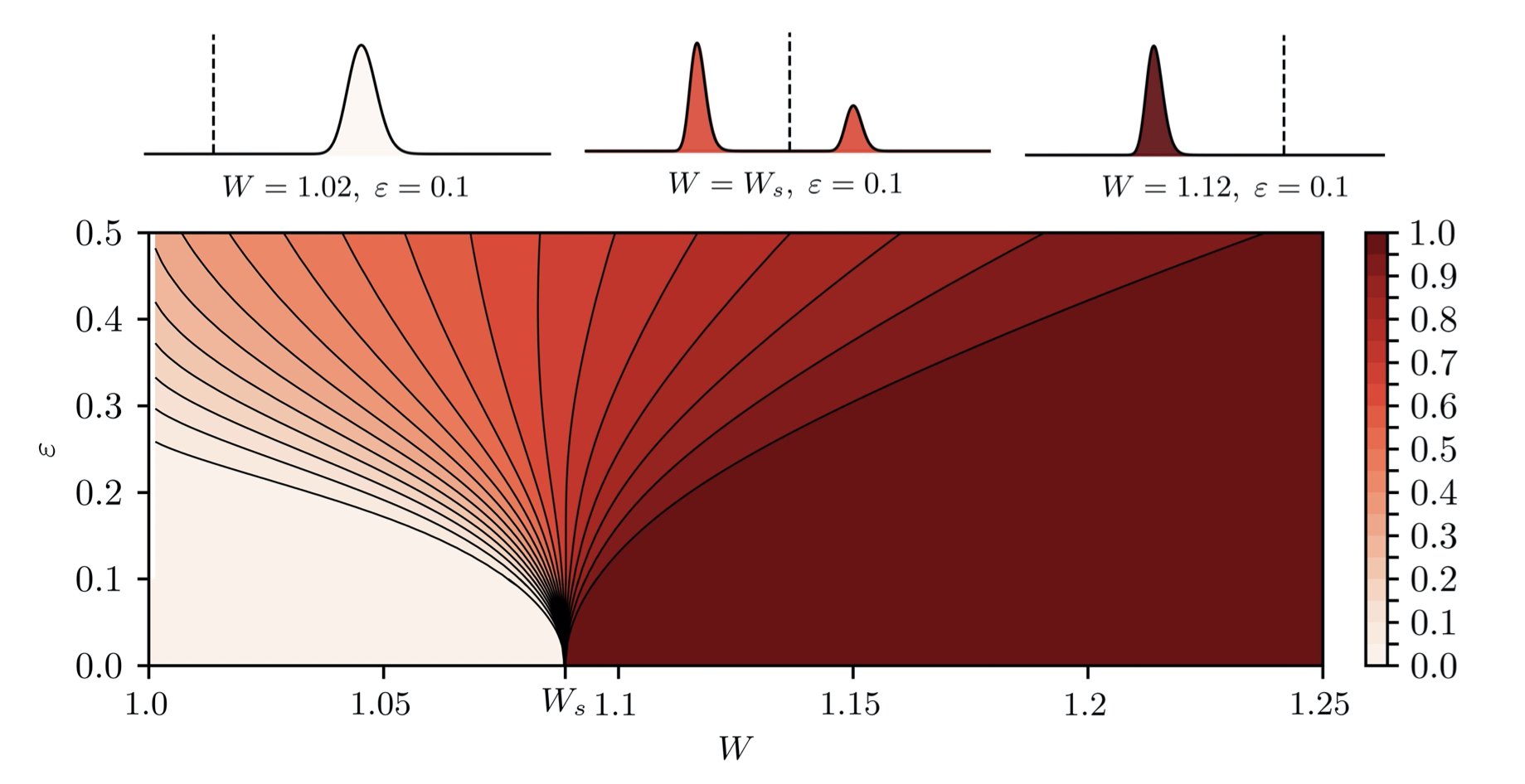}
    \caption{The steady-state probability
      $\mathbbm{P}\{B_{t}^{*}\in \mathscr{B}_{1}\}$ of finding the
      state in the basin of attraction for reverse flow as a
      function of wind strength $W$ and fluctuation strength $\varepsilon$ when
      $\sigma(b;W)=\varepsilon b$. At the top of the figure are
      shown stationary probability densities for $\varepsilon=0.1$
      at various wind strengths, for which the dashed line
      indicates the threshold $b=W$ separating forward flow ($b>W$) from
      reverse flow ($b<W$).}
    \label{fig:forward_reverse}
\end{figure}

For practical applications it is useful to identify combinations
of $W$ and $\varepsilon$ that lead to confidence in whether the
statistically steady state corresponds to forward or reverse
flow. To this end, figure \ref{fig:forward_reverse} plots isolines
of the stationary probability
$\mathbbm{P}\{B_{t}^{*}\in \mathscr{B}_{1}\}$, which is the
probability of finding the system in the basin of attraction for
reverse flow. Because the line $b=W$ demarcating forward and
reverse flow (see, for example, figure \ref{fig:stationary}) does
not coincide with the local maxima of $V_{B}$, reverse flow is
theoretically possible even when the system is in the basin of
attraction for forward flow. Consequently, probabilities
corresponding to the likelihood of being in the basin of attraction
for reverse flow underestimate (by an amount that diminishes with
$\varepsilon$) the probability of forward flow, i.e.
$\mathbbm{P}\{B_{t}^{*}\in \mathscr{B}_{1}\}<\mathbbm{P}\{B_{t}^{*}<W\}$.

The contours in figure \ref{fig:forward_reverse} are characterised
by a fan shaped region of uncertainty, in which the probability
varies continuously between $0$ and $1$, emanating from
$(W,\varepsilon)=(W_{s},0)$. On either side of this region of
uncertainty are regions for which one can be close to certain that
the system will eventually be in either the basin of attraction
for forward or reverse flow. Plots such as
\eqref{fig:forward_reverse} might provide a useful guide to engineers by
identifying conditions for which it would be difficult or
inappropriate to design for a single flow regime.

\begin{figure}[t]
    \centering
    \includegraphics{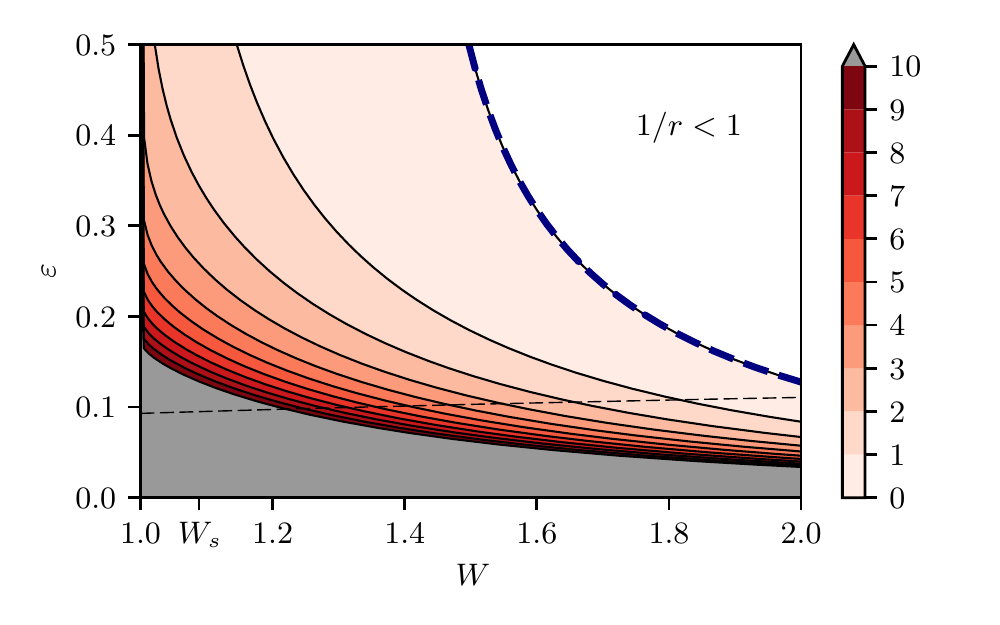}
    \caption{The logarithm of the inverse of Kramer's escape rate
      $\log(1/r)$ from \eqref{eq:kramers} as a function of the wind strength $W$ and
      fluctuation strength $\varepsilon$ when $\sigma(b;W)=\varepsilon b$. The blue dashed line is
      the contour for which $r=1$. The black dashed line corresponds to \eqref{eq:dim_epsilon3} in \S\ref{sec:app} for $\hat{S}=50\,\mathrm{m}^{2}$, $\hat{H}=3\,\mathrm{m}$, $\hat{A}=0.5\,\mathrm{m}^{2}$ and $\lambda = 1$.}
    \label{fig:escape}
\end{figure}

\subsection{Kramer's escape rate}
\label{sec:kramer}

The stationary densities discussed in the previous section are
solutions to $\partial_{t}f_{B}=0$. How long one would have to
wait in practice to see such solutions, however, would depend on
the initial state of the system. For example, for small values of
$\varepsilon$, wind strengths $W$ slightly larger than $W_{s}$ and
an initial density in the region $\mathscr{B}_{3}$ (forward flow),
one might have to wait a long time for the density to be transported into
$\mathscr{B}_{1}$ (reverse flow) in the way that figure
\ref{fig:forward_reverse} suggests. The physical reason for such
large time scales is that when $W\approx W_{s}$ the potential
wells are both relatively deep, which, for small
$\varepsilon$, means that a rare (infrequent) fluctuation in the
system's forcing is required to produce a transition between the
two wells.

The remarks above can be made precise by considering a potential
barrier $\Delta V_{X}$, quantified as the difference between the
local maxima $V_{X}(x_{2})$ and local minima $V_{X}(x_{3})<V_{X}(x_{2})$
associated with forward flow,

\begin{equation}
  \Delta V_{X}:= V_{X}(x_{2}(W))-V_{X}(x_{3}(W))\geq 0\quad \forall W\geq 1.
\end{equation}

\noindent If the potential barrier is large relative to the
diffusivity $\varepsilon$, the rate $r(W,\varepsilon)$ at which
states `escape' from $\mathscr{B}_{3}$ (forward flow), quantified
as the probability flux divided by the probability, can be
approximated using Kramer's escape rate formula
\cite{RisHboo1996a}:

\begin{equation}
r(W,\varepsilon) = \frac{1}{2\pi}\sqrt{V_{X}''(x_{3})\lvert V_{X}''(x_{2})\rvert}\,\mathrm{exp}\left(-\frac{2\Delta V_{X}}{\varepsilon^{2}} \right).
\label{eq:kramers}
\end{equation}

\noindent The associated timescale, $1/r$, is plotted in figure
\ref{fig:escape} for the quadratic diffusion model
$\sigma(b;W)=\varepsilon b$. The figure shows that the time one
can expect to wait to reach a stationary state decreases with
respect to $W$, which controls the height of the potential barrier,
and increases with respect to $\varepsilon$, which quantifies the
`destabilising' noise. Since $V_{X}$ does not depend on
$\varepsilon$, \eqref{eq:kramers} suggests that the time $1/r$
that one would have to wait to reach a stationary density increases
exponentially with respect to $1/\varepsilon^{2}$.

\begin{figure}[t]
    \centering
    \includegraphics[width=\textwidth]{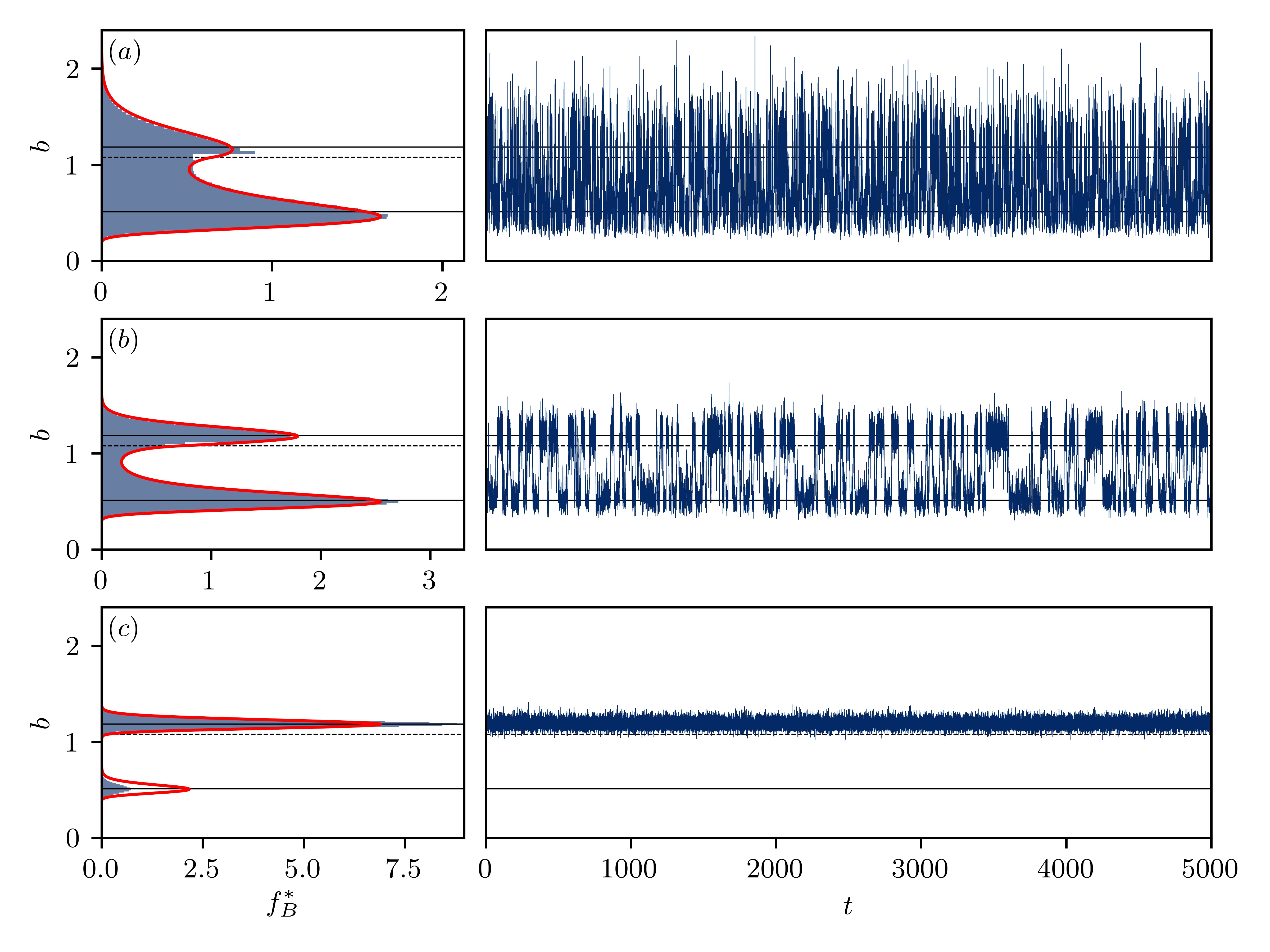}
    \caption{Realisations of the system using \eqref{eq:euler} for
      $W=W_s-0.01$, $\varepsilon =0.5\ (a)$,
      $\varepsilon =0.25\ (b)$ and $\varepsilon =0.125\ (c)$. The
      panels on the left show the stationary probability density
      $f^{*}_B$ collected from simulation data (blue filled
      region) and compared with the analytical solution
      \eqref{eq:fB} (red line), whereas the panels on the right
      show a single realisation of buoyancy $B_{t}$. The dashed
      line corresponds to $b=W$ and therefore indicates the
      threshold separating forward flow ($b>W$) from reverse flow
      $(b<W)$. The horizontal black lines denote the location of
      the fixed points for forward flow and reverse flow.}
    \label{fig:Ws}
\end{figure}

\section{Numerical simulations}\label{sec:simulations}

To obtain individual realisations of the stochastic model,
\eqref{eq:sde} can be integrated directly. Using the Heun method
\cite{BurKprs2003a}, which accounts for the drift
induced by the Stratonovich interpretation of the noise without
modification of the function `$a$' in \eqref{eq:sde}, the discretised equations are

\begin{equation}
  B_{t_{n+1}} = B_{t_{n}}+\overline{a}(B_{t_{n}};W)\Delta t_{n} + \overline{\sigma}(B_{t_{n}}; W)\Delta \xi_{t_{n}},
\label{eq:euler}
\end{equation}

\noindent where $\Delta t_{n}:=t_{n+1}-t_{n}$, $\Delta\xi_{t_{n}}:=\xi_{t_{n+1}}-\xi_{t_{n}}\sim N(0,\Delta t_{n})$ and 
\begin{align}
\overline{a}(B_{t_{n}};W)&:=\frac{1}{2}\left[a(\tilde{B}_{t_{n+1}};W)+a(B_{t_{n}};W)\right],\\
\overline{\sigma}(B_{t_{n}};W)&:=\frac{1}{2}\left[\sigma(\tilde{B}_{t_{n+1}};W)+\sigma(B_{t_{n}};W)\right],
\label{eq:euler_av}
\end{align}
are intermediate values of the drift and standard deviation based
on the following estimate of the buoyancy at time $t_{n+1}$:

\begin{equation}
\tilde{B}_{t_{n+1}}:=B_{t_{n}}+a(B_{t_{n}};W)\Delta t_{n}+\sigma(B_{t_{n}}; W)\Delta \xi_{t_{n}}.
\label{eq:euler_mid}
\end{equation}

\begin{figure}[t]
    \centering
    \includegraphics[width =\textwidth]{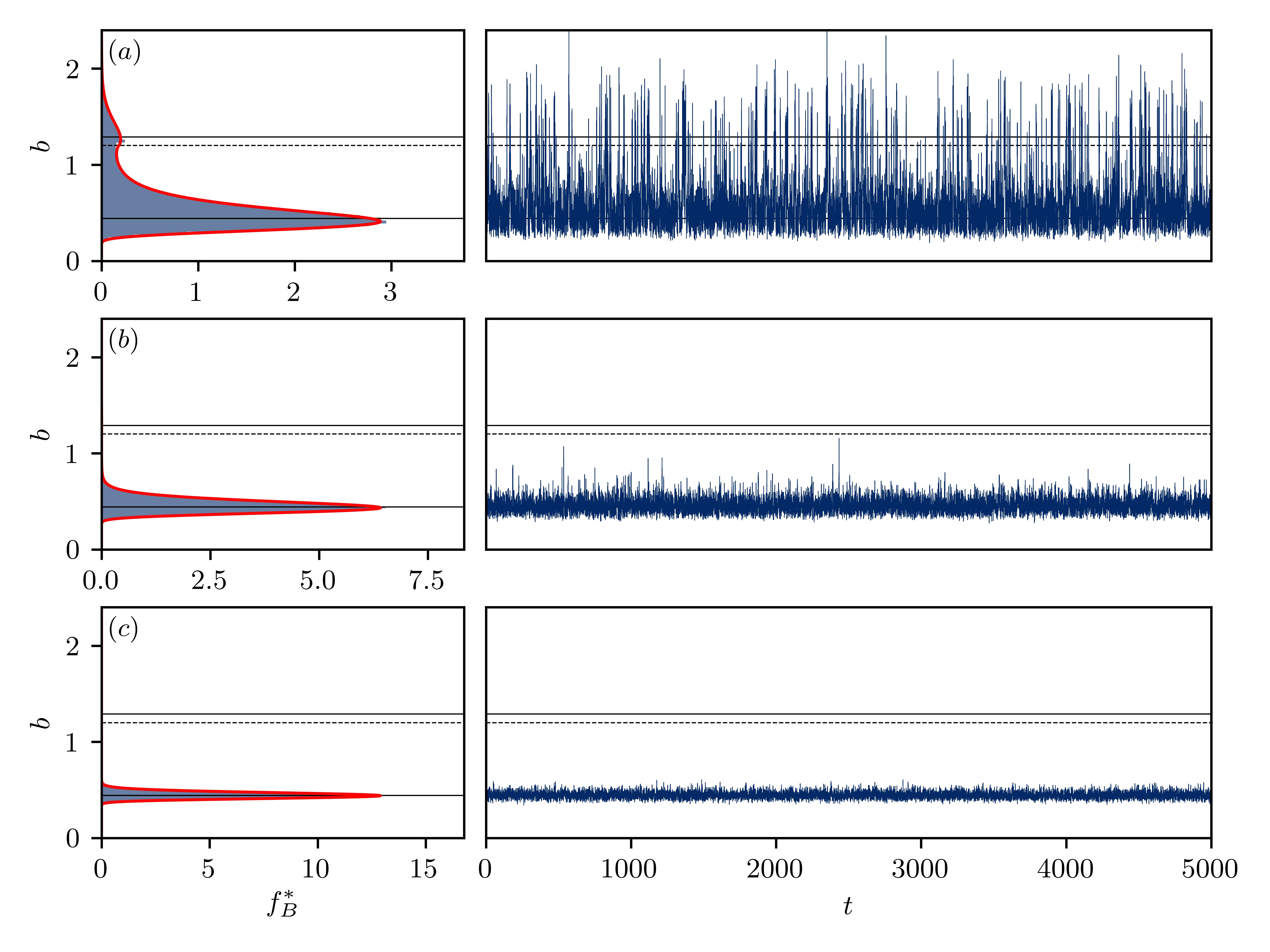}
    \caption{Realisations of the system using \eqref{eq:euler} for
      $W=1.2$, and $\varepsilon =0.5\ (a)$,
      $\varepsilon =0.25\ (b)$ and $\varepsilon =0.125\ (c)$. The panels on the left show the
      stationary probability density $f^{*}_B$ collected from simulation data
      (blue filled region) and compared with the analytical
      solution \eqref{eq:fB} (red line), whereas the panels on the
      right show a single realisation of buoyancy $B_{t}$. The dashed
      line corresponds to $b=W$ and therefore indicates the
      threshold separating forward flow ($b>W$) from reverse flow
      $(b<W)$. The horizontal black lines denote the location of the fixed points for forward flow and reverse flow.}
    \label{fig:W12}
\end{figure}

The sample trajectories shown in figures \ref{fig:Ws} and
\ref{fig:W12} were obtained with $\Delta t_{n}=0.05$ for
$t_{n} \in [0,5000]$.  A set of $300$ such realisations were
obtained for a given value of $W$ to construct the histograms
shown on the left-hand side of figures \ref{fig:Ws} and
\ref{fig:W12} (in this regard, note that each time series on the
right-hand side is a single realisation). The stable fixed point for either
forward flow ($W<W_{s}$ shown in figure \ref{fig:Ws}) or reverse
flow ($W>W_{s}$ shown in figure \ref{fig:W12}) was used as an
initial condition for all realisations and statistically transient
data for $t_{n} \in [0,100]$ were not included in the construction
of the histograms.

Figure \ref{fig:Ws} shows realisations for $W=W_{s}-0.01$
(cf. figure \ref{fig:approximation}, for which the probability
density would concentrate itself around the fixed point for
forward flow in the limit $\varepsilon\rightarrow 0$). For cases
$\varepsilon=0.5$ and $\varepsilon=0.25$, illustrated in the
subplots $(a)$ and $(b)$, respectively, the system's trajectory
crosses the potential barrier frequently, which leads to large
variance and significant uncertainty regarding the direction of
the ventilation. For $\varepsilon=0.125$ in subplot $(c)$,
however, the system is likely to be found in forward flow and does
not appear to transition between forward and reverse flow
frequently. In contrast, figure \ref{fig:W12} shows that when
$W=1.2$ the system resides predominantly in reverse flow for the
reasons described in \S\ref{sec:solutions}. Whereas
$\varepsilon=0.5$ produces occasional excursions of the system
into a state of forward flow, for $\varepsilon=0.25$ and
$\varepsilon=0.125$ the flow is almost entirely in the reverse
direction.

The histograms shown in figures \ref{fig:Ws} and \ref{fig:W12}
exhibit a reasonably good agreement with the analytical solution
\eqref{eq:fB}. The differences between the analytical solution and
the histograms obtained from realisations of \eqref{eq:sde} in
figure \ref{fig:Ws}$(c)$ illustrate the challenge of obtaining a
sufficient number of samples for a system that undergoes
infrequent transitions. Bearing in mind that the governing
equations are nondimensional, figure \ref{fig:Ws} shows that one
needs to collect data over extremely long durations/large
ensembles in order to resolve the underlying density, which makes
the availability of analytical solutions or approximations
particularly attractive.

\begin{figure}[t]
    \centering
    \includegraphics{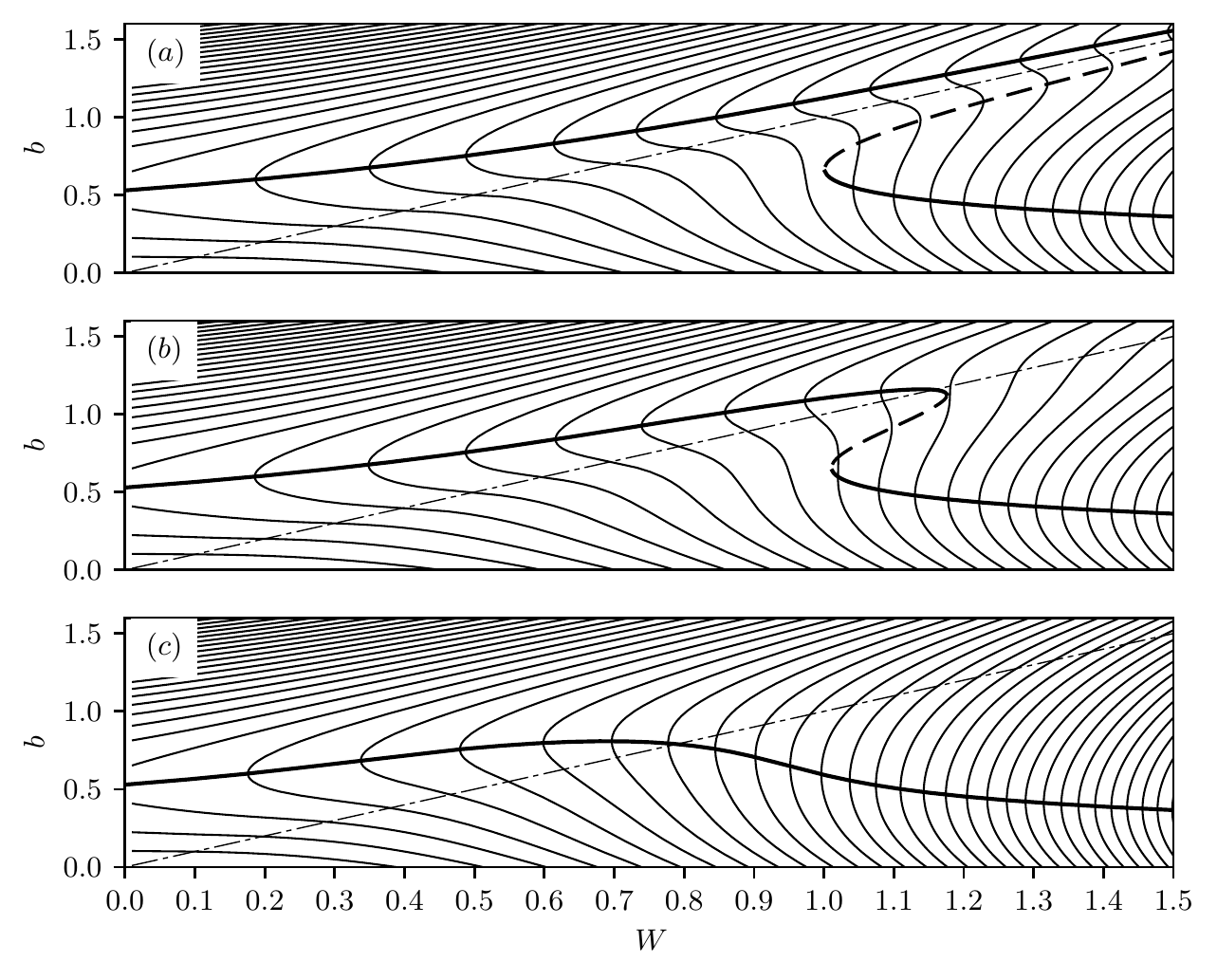}
    \caption{Contours of the modified potential $\tilde{V}_{B}$ to account
      for stochastic wind using an adiabatic approximation
      for $\gamma/\nu^{2}=1000$ $(a)$, $\gamma/\nu^{2}=100$ $(b)$
      and $\gamma/\nu^{2}=10$ $(c)$. The thick line corresponds to
      (stable) local minima of the potential while the thick
      dashed line corresponds to (unstable) local maxima of the
      potential. The dash-dot line corresponds to $b=W$ and
      therefore demarcates forward flow from reverse flow.}
    \label{fig:bifurcations}
\end{figure}

\section{An Ornstein-Uhlenbeck (`Langevin') model}
\label{sec:orn}

As discussed in \S\ref{sec:stochastic}, the effects of
fluctuations can be incorporated in several ways, which raises
important questions concerning their physical origin and the
correct interpretation of the deterministic equations to which
they are added. In \S\ref{sec:stochastic} fluctuations were
included as a stochastic buoyancy flux; in this section, the
stochastic evolution of the wind is specified as an additional
equation. In general, such an equation could be adapted to produce
a specific wind speed distribution from a given set of
environmental conditions \cite{MaJene2018a}. One particular, if
somewhat idealised, approach is to represent fluctuations in the wind
velocity that affect the bulk pressure difference
($\Delta \hat{p}$ in \S\ref{sec:det}) as an Ornstein-Uhlenbeck
process \cite{VesRbae2023a}, which can account for temporal
correlations. If the stochastic buoyancy flux from \S\ref{sec:det}
is retained, the governing equations become

  \begin{align}
    \rd B_{t} &= a(B_{t},U_{t};W)\rd t + \sigma(B_{t}; W)\circ \rd \xi_{t}, \label{eq:sde2}\\
    \rd U_{t} &= -\gamma U_{t}\rd t + \nu\rd \zeta_{t},\label{eq:ou}
  \end{align} 

\noindent where, following \cite{VesRbae2023a}, $U_{t}$
corresponds to dimensionless fluctuations in the wind velocity, which is
proportional to the square root of the resulting pressure difference:
 
\begin{equation}
  a(B_{t},U_{t};W):=1-c\lvert B_{t}-(1+U_{t})^{2}W\rvert^{1/2}\,B_{t}.
\end{equation}

\noindent The parameter $\gamma$ determines the rate at which
$U_{t}$ reverts to its mean value of zero and $\nu\rd \zeta_{t}$
denotes uncorrelated increments of a zero-mean Gaussian process
with time- and state-invariant standard deviation $\nu$. One difference between
\eqref{eq:sde2}-\eqref{eq:ou} and the simpler version studied in
\S\ref{sec:stochastic} is that the stochastic forcing appears in
the nonlinear drift term $a(B_{t},U_{t};W)$. A second difference is that the
probability density corresponding to \eqref{eq:sde2}-\eqref{eq:ou}
is a joint density $f_{BU}$ of two (rather than one) variables.

The Fokker-Planck equation for the joint density $f_{BU}$ is

\begin{equation}
  \pd{f_{BU}}{t}=-\nabla\cdot \vc{J}_{BU},
  \label{eq:fBU}
\end{equation}

\noindent where

\begin{equation}
  \renewcommand*{\arraystretch}{1.3}
  \vc{J}_{BU}:=
  \begin{pmatrix}
    1-\sqrt{|b-(1+u)^{2}W|}b-\frac{1}{2}\sigma\partial_{b}\sigma \\
    -\gamma u -\frac{1}{2}\nu^{2}\partial_{u}
  \end{pmatrix}
  f_{BU},
  \label{eq:JBU}
\end{equation}

\noindent is the two-dimensional probability flux. Integrating
\eqref{eq:fBU} with respect to $u\in (-\infty,\infty)$, gives an
equation for $f_{B}$ that is satisfied by the stationary marginal density $f_{B}^{*}$, which is the
quantity of interest:

\begin{equation}
  \pd{}{b}\left(\mathbbm{E}[a(B_{t},U_{t}|W)|B_{t}=b]f_{B}\right)-\frac{1}{2}\pd{}{b}\left(\sigma\left(\pd{}{b}\sigma f_{B}\right)\right)=0,
\label{eq:fB2}
\end{equation}

\noindent where, making use of the conditional density $f_{U|B}:=f_{BU}/f_{B}$ when $f_{B}\neq 0$,

\begin{equation}
  \mathbbm{E}[a(B_{t},U_{t}|W)|B_{t}=b]:=\int\limits_{-\infty}^{\infty}a(b,u|W)f_{U|B}(u, t;b)\rd u,
\label{eq:Ea}
\end{equation}

\noindent is the unknown conditional expectation of the the drift
$a(B_{t},U_{t}|W)$ for a given value $b$ of $B_{t}$. Unless the full
equation \eqref{eq:fBU} is solved, an assumption about the
conditional expectation is required to make progress with
\eqref{eq:fB2}.

If $\gamma$ is sufficiently large, the modelled wind will evolve
on a much faster time scale than the buoyancy and can be regarded
as being in a state of quasi-equilibrium. Corresponding to such a
separation of time scales is the assumption that derivatives with
respect to $u$ of the conditional density $f_{U|B}$ are much
larger than those with respect to $b$. Consequently, when
$f_{BU}=f_{U|B}f_{B}$ is substituted into \eqref{eq:fBU} it is
seen that $f_{U|B}$ can be approximated by a balance of terms
involving only relatively large derivatives with respect to $u$ \cite{HakHboo2013a}:

\begin{equation}
 \gamma \pd{}{u}(uf_{U|B}) - \frac{\nu^{2}}{2}\frac{\partial^{2}}{\partial u^{2}}f_{U|B}\sim 0,
\end{equation}

\noindent which has the well-known solution \cite{GarCboo1989a}

\begin{equation}
  f^{*}_{U|B}(u;b)\sim \sqrt{\frac{\gamma}{\pi \nu^{2}}}\exp\left(-\frac{\gamma u^{2}}{\nu^{2}}\right).
  \label{eq:fU}
\end{equation}

\noindent In this particular case, for which $B_{t}$ does not
feature in \eqref{eq:ou}, $f^{*}_{U|B}$ in \eqref{eq:fU} corresponds
to the stationary marginal density $f^{*}_{U}$.

The subsequent use of $f_{U|B}^{*}$ obtained from \eqref{eq:fU} in
\eqref{eq:Ea} to account for the effect that the fast variable has
on the evolution of the slow variable, while removing the former
from the calculations, is known as `adiabatic elimination'
\cite{TheWpha1985a, HakHboo2013a}. With such an approximation, the
drift term in \eqref{eq:fB2} can be expressed as the gradient of
the potential \eqref{eq:potential} that is convolved with the
marginal density of the fast variable:

\begin{equation}
  \pd{}{b}\left(-\od{\tilde{V}_{B}}{b}f_{B}-\frac{1}{2}\sigma\pd{}{b}\left(\sigma f_{B}\right)\right) =0,
\label{eq:fB3}
\end{equation}

\noindent where, noting that $f^{*}_{U|B}=f^{*}_{U}$ from \eqref{eq:fU} does not depend on $b$,

\begin{equation}
  \tilde{V}_{B}\left(b;\frac{\gamma}{\nu^{2}}\right):=\int\limits_{-\infty}^{\infty}V_{B}\left(b; (1+u)^{2}W\right)f^{*}_{U}(u)\rd u.
  \label{eq:Vs}
\end{equation}

Figure \ref{fig:bifurcations} displays contours of $\tilde{V}_{B}$
along with `equilibrium' paths corresponding to its local
extrema. Sufficiently large wind fluctuations (small
$\gamma/\nu^{2}$) modify the original potential (see cf. figure
\ref{fig:potential}) via the convolution \eqref{eq:Vs} by removing
the metastable equilibrium corresponding to forward flow at large
wind strengths $W$.  When $\gamma/\nu^{2}=10$, the potential is
convex and has a unique minimum corresponding to forward or
reverse flow for small or large values of $W$, respectively. It is
in this way that the nonlinear effects of fluctuations in the wind
strength in \eqref{eq:sde2} modify the underlying structure of the
equation, regardless of noise in the buoyancy flux that can be
included separately in \eqref{eq:sde2} with a particular parameterisation for $\sigma$.

\begin{figure}[t]
    \centering
    \includegraphics{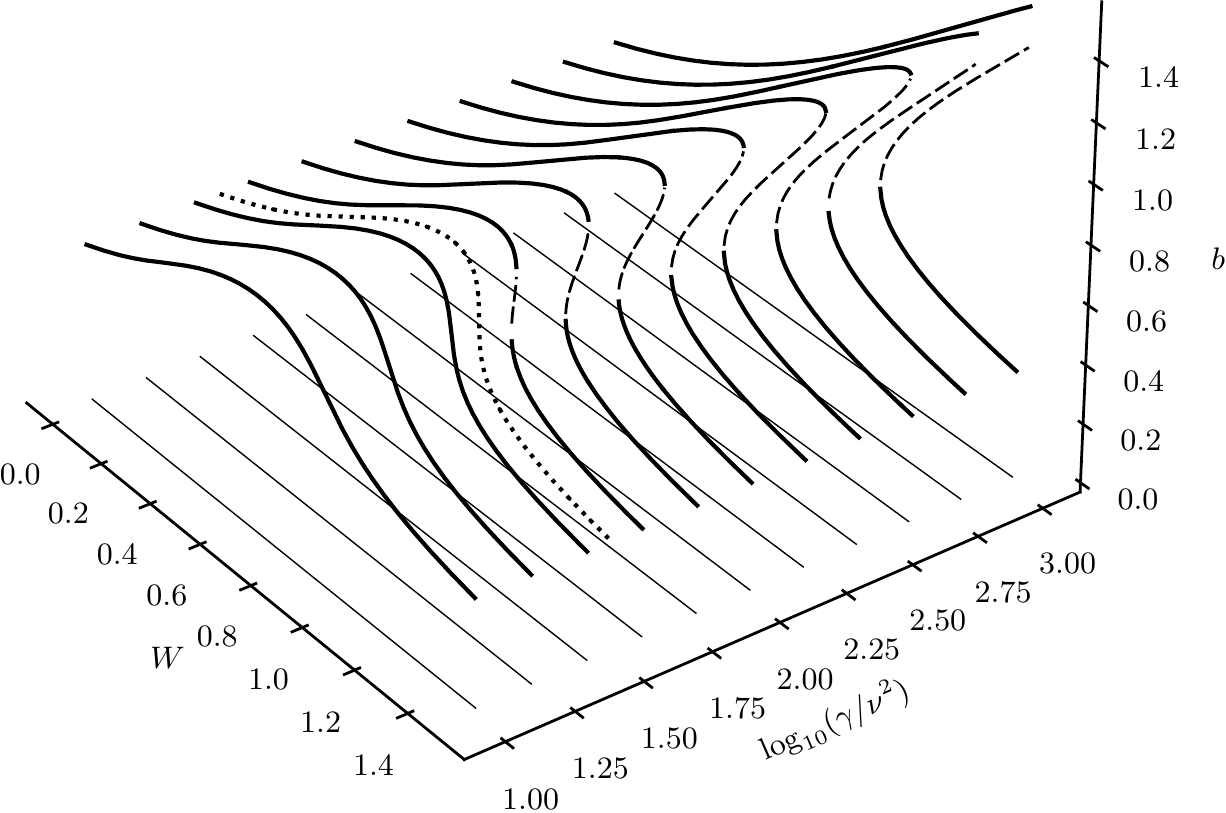}
    \caption{The equilibrium surface of the modified potential $\tilde{V}_{B}$ derived in
      \S\ref{sec:orn} as a function of the base wind strength $W$, buoyancy $b$ and the
      parameters of the Ornstein-Uhlenbeck process. The dotted
      line corresponds to $\gamma/\nu^{2}=30.775$, which is the
      value above which the modified potential admits an unstable
      fixed point and an additional stable fixed point
      corresponding to forward flow.}
    \label{fig:catastrophe}
\end{figure}

As illustrated in figure \ref{fig:catastrophe}, which shows
equilibria of the potential surface $\tilde{V}_{B}$ as a function of
both $W$ and $\gamma/\nu^{2}$, the cusp point of the potential,
above which a second local minimum in $\tilde{V}_{B}$ emerges, is
$\gamma/\nu^{2}=30.775$. The relevance of this number to practical
applications will be discussed at the end of the next section.

\begin{figure}
    \centering
    \includegraphics[width=\textwidth]{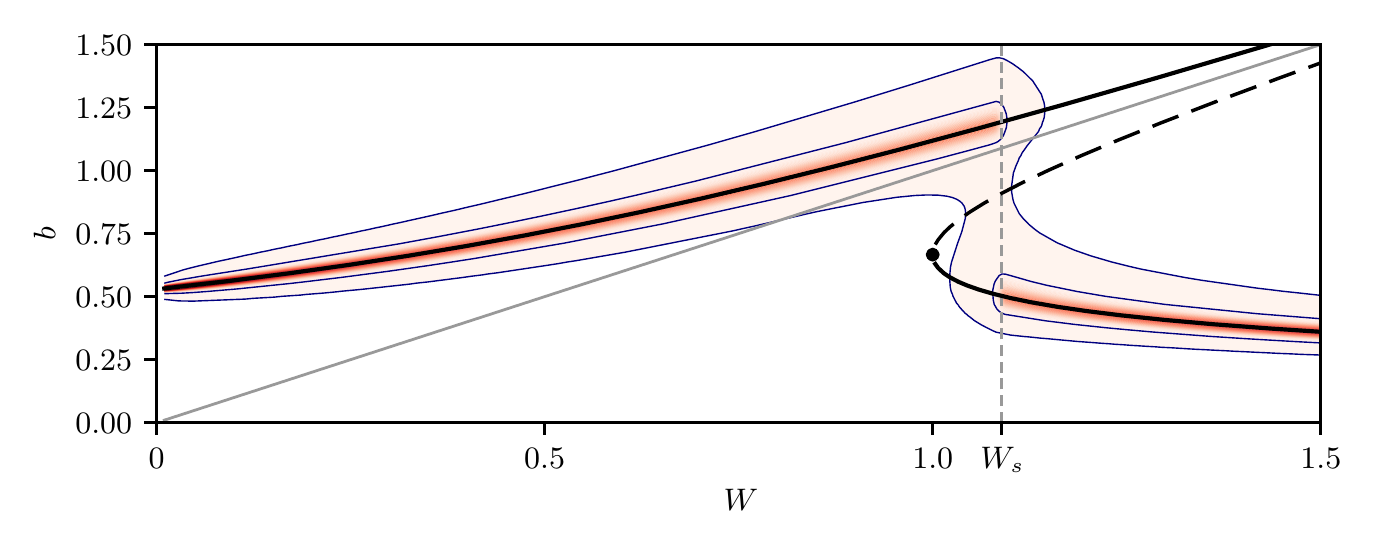}
    \caption{The stationary probability density $f^{*}_{B}$ (shades of red) as
      a function of the wind strength $W$ when
      $\sigma(b;W)=\varepsilon b$ for
      $\varepsilon\propto W^{1/4}$, as defined by
      \eqref{eq:dim_epsilon3} with $\hat{S}=50\,\mathrm{m}^{2}$,
      $\hat{H}=3\,\mathrm{m}$, $\hat{A}=0.5\,\mathrm{m}^{2}$ and
      $\lambda = 1$. The thin blue lines highlight contours
      corresponding to a stationary density of $10^{-8}$ and
      $5\times 10^{-1}$.}
    \label{fig:stationary_nonuniform}
\end{figure}

\section{Application}
\label{sec:app}

Although it is convenient to use nondimensionalised equations
\eqref{eq:det}, it is useful to consider how the results relate to
the dimensional parameters that might be encountered in
practice. For the quadratic diffusion model applied to the
buoyancy flux in \S\ref{sec:quad}, $\varepsilon$ characterises the
effects of uni- or bi-directional fluctuations in the flow through
an opening. While it is therefore difficult to relate wind speed
statistics to $\varepsilon$ in a precise way without a model
developed from first principles, it is instructive to consider the
dimensional version of the equations (cf. \eqref{eq:det_dim}), for
which care is needed in deducing the correct nondimensionalisation
of $\varepsilon$:

\begin{equation}
  \hat{H}\hat{S}\rd \hat{B}_{t}=(\hat{F}-\hat{Q}\hat{B}_{t})\rd \hat{t} + \hat{\sigma}\rd\hat{\xi}_{t}.
  \label{eq:dim_sigma}
\end{equation}

\noindent In \eqref{eq:dim_sigma} the Wiener process has
dimensions corresponding to the square root of time and
$\hat{\sigma}:=\hat{\varepsilon}\hat{B}_{t}$ is the dimensional
version of \eqref{eq:sigma}. Therefore, noting that each term in
\eqref{eq:dim_sigma} has dimensions $L^{4}T^{-2}$, where
the dimensions of length and time are denoted as $L$ and $T$, respectively,
implies that $[\hat{\varepsilon}]=L^{3}T^{-1/2}$.

If $\hat{v}$ corresponds to a velocity scale that characterises
fluctuations in the buoyancy flux through the opening, then it is
reasonable to express $\hat{\varepsilon}$ in terms of $\hat{v}$
and the opening area $\hat{A}$, such that
$\hat{\varepsilon}=\hat{v}^{1/2}\hat{A}^{5/4}$ on dimensional
grounds; hence, nondimensionalising $\hat{B}_{t}$ and
$\hat{\xi}_{t}$ in \eqref{eq:dim_sigma} using
\specialeqref{eq:dims}{a} and \specialeqref{eq:dims}{b},
respectively,

\begin{equation}
  \rd B_{t}=\ldots+\frac{\hat{\varepsilon}}{\hat{H}\hat{S}}B_{t}\rd\hat{\xi}_{t}= \ldots +\frac{\hat{v}^{1/2}\hat{A}^{5/4}}{\hat{H}\hat{S}}\left({\frac{27\hat{H}^{2}\hat{S}^{3}}{4\hat{A}^{2}\hat{F}}}\right)^{1/6}B_{t}\rd\xi_{t},
  \label{eq:dim_epsilon}
\end{equation}

\noindent which, when compared with \eqref{eq:sde}, shows that

\begin{equation}
  \varepsilon = \left(\frac{27^{2}\hat{v}^{6}\hat{A}^{11}}{4^{2}\hat{H}^{8}\hat{S}^{6}\hat{F}^{2}}\right)^{1/12}.
  \label{eq:dim_epsilon2}
\end{equation}

\noindent If the strength of the fluctuations is related to the
wind-induced pressure difference by
$\lambda \Delta\hat{p}/\hat{\rho}_{0}=\hat{v}^{2}/2$, where
$\lambda$ is a dimensionless number that can account for
orientation and the statistical properties of the wind, then
\eqref{eq:pc} and \specialeqref{eq:dims}{c} can be used to express
$\hat{F}$ in terms of $W$, such that \eqref{eq:dim_epsilon2} becomes

\begin{equation}
  \varepsilon = \left(\frac{27\hat{A}^{3}\,\lambda W}{2\hat{H}^{2}\hat{S}^{2}}\right)^{1/4}.
  \label{eq:dim_epsilon3}
\end{equation}

Equation \eqref{eq:dim_epsilon3} is useful because it provides an
algebraic relationship between $\varepsilon$ and $\lambda W$ for
which the prefactor is determined by the geometric properties of
the room/building. For a typical meeting room of cross-sectional
area $\hat{S}=50\,\mathrm{m}^{2}$, height $\hat{H}=3\,\mathrm{m}$
and effective opening area $\hat{A}=0.5\,\mathrm{m}^{2}$,
\eqref{eq:dim_epsilon3} implies that, for $W=1$ (corresponding to
$\hat{F}=0.18\,\mathrm{m}^{4}\mathrm{s}^{-3}$, which is equivalent
to approximately $7\,\mathrm{kW}$ of heat), $\varepsilon=0.093$.
More generally, figure \ref{fig:stationary_nonuniform} displays
the stationary probability density using \eqref{eq:dim_epsilon3}
to relate $\varepsilon$ and $W$ for $W\in [0,1.5]$, $\lambda=1$
and the geometric parameters stated above. The central message
conveyed by figure \ref{fig:stationary_nonuniform} is that, for
realistic geometric parameters and wind fluctuations that are of
the same order of magnitude as the strength of an opposing wind,
sustained buoyancy-driven flow opposing the wind direction is
extremely unlikely for wind strengths that exceed the
statistically critical value $W_{s}$. However, for statistically
transient states arising, for example, from temporal changes in
either $\lambda$ or $W$, characterising the behaviour of the
system probabilistically is more challenging.

The time scale associated with the dimensional parameters
discussed above is 

\begin{equation}
  \frac{\hat{t}}{t}:=\left(\frac{27\hat{H}^{2}\hat{S}^{3}}{4\hat{A}^{2}\hat{F}}\right)^{1/3}\approx 551\,\mathrm{s}.
\label{eq:t_hat}
\end{equation}

\noindent Multiplying the nondimensional time scale from figure
\ref{fig:escape} by \eqref{eq:t_hat} along the dashed line
corresponding to \eqref{eq:dim_epsilon3} shows that the time one
would have to wait for such a system to settle into the stationary
state shown in figure \ref{fig:stationary_nonuniform}, having
started in forward flow, could be comparable to the durations for
which such a room is likely to be used (i.e. in the order of one
hour). Considerations of the transient evolution of the
probability density would therefore play an important role in such
applications.  While relatively small fluctuations correspond to
certainty regarding the eventual state of the flow, they
correspond to uncertainty in producing a longer transient period
than relatively large fluctuations. However, for relatively large
base wind strengths (e.g. $W\geq 2$), it is likely that the system
would settle into reverse flow in a matter of minutes.

For the Ornstein-Uhlenbeck process discussed in \S\ref{sec:orn},
the random variable $U_{t}$ corresponds to fluctuations in the
wind speed divided by the base velocity scale
$\sqrt{2\Delta\hat{p}/\hat{\rho}_{0}}$ used to define $W$ in
\eqref{eq:dims}. The statistics reported in a recent study on
stochastic models for wind\cite{MaJene2018a, VesRbae2023a}
suggest that plausible values of $\gamma/\nu^{2}$ in \eqref{eq:fU}
lie between $10$ and $1000$ and therefore encompass the
bifurcation illustrated in figure
\ref{fig:catastrophe}. Relatively small values of $\gamma/\nu^{2}$
within this range (relatively long, large gusts of wind) will lead
to reverse flow for large base wind strengths, whereas relatively
large values of $\gamma/\nu^{2}$ (relatively short, small gusts of
wind) do not necessarily rule out the possibility of forward flow
at large base wind strengths.

\section{Conclusions}
\label{sec:conclusions}

This work has described stochastic models for natural ventilation
driven by opposing wind and buoyancy. The results expose several
properties concerning the system's predictability and stability
for the first time and highlight challenges associated with the
incorporation of stochastic effects into existing deterministic
models for ventilation.

In the absence of rigorous justification for a particular form of
stochastic forcing from either theory or experiments, the models
chosen for this study were simple and of general applicability. In
\S\ref{sec:stochastic} a stochastic buoyancy flux was added to the
deterministic evolution of buoyancy in the form of a Stratonovich integral of
a Wiener process \eqref{eq:sde} as the limit of a stochastic
process with relatively short autocorrelation time. In general,
the approach can account for fluctuations of the buoyancy flux
that depend on the system's state and the base wind strength
$W$. The evolution of the resulting probability density is
determined by a Fokker-Planck equation which accounts for the
stabilising effects of the underlying potential and the
destabilising effects of noise. An analytically-defined
Boltzmann-like distribution describes the corresponding stationary
density. In \S\ref{sec:orn} the wind itself was modelled as
an Ornstein-Uhlenbeck process at the expense of introducing an
additional dimension, which was subsequently approximated using
adiabatic elimination.

Although the deterministic system for this problem admits multiple
steady states, for sufficiently large opposing wind strengths and
small amounts of noise, the system's stationary state is unlikely
to be found in the metastable region associated with forward
flow. As the wind strength increases, fluctuations in the buoyancy
flux are, with increasing likelihood, able to overcome the
relatively small potential barrier separating forward and reverse
flow. Another reason is that when the wind itself is modelled as a
stochastic process (see \S\ref{sec:orn}), fluctuations in the wind
convolve with the nonlinear function describing the system's
underlying potential, which can lead to the removal of the
metastable equilibrium altogether, as shown in figure
\ref{fig:bifurcations}. In these respects, the state of the
stochastic system, which accounts for disturbances that are likely
to be present in practical applications, is easier to `determine'
in a design context than it would be for the idealised
`deterministic' system. Although the particularities of such
results depend on the choice of stochastic model, their
qualitative properties are generic and expected to be found in a
broad range of possible stochastic models.

An important caveat to the conclusions above is that certainty in
the eventual state of the system goes hand in hand with
uncertainty associated with an increase in the duration of the
transient evolution of the probability density. It is therefore
difficult to be prescriptive about whether the system is likely to
be in a particular state without careful consideration of the base
wind strength and relative size of the fluctuations. Base wind
strengths that are close to critical are the most difficult in
this regard, because they are associated with bimodal probability
densities with transients of the longest duration.

A preeminent challenge to the future incorporation of stochastic
effects in existing deterministic models of ventilation is finding
physically faithful interpretations and representations of
fluctuations, and reconciling them with the established
deterministic part of such models. In bulk models of turbulence it
is not always clear whether the established deterministic
equations represent the limit of a system with zero external noise
or as the macroscopic average of a system with finite (internal)
noise. The former interpretation is somewhat easier to approach
theoretically and, if regarded as the limit of a process with
relatively short autocorrelation times, is consistent with our use
of the Stratonovich integral in \eqref{eq:sde}
\cite{KamNjsp1981a}.

When systems involve internal noise (which is arguably always the
case where turbulence is concerned) the situation is more
complicated because the fluctuations can't be removed. In that
case it would be necessary to return to fundamentals and develop a
master equation for the system's probability density from first
principles, rather than trying to ascribe meaning to an otherwise
ill-defined deterministic equation \eqref{eq:det} from a Langevin
perspective \cite{KamNjsp1981a}. With a suitable master equation
determining every aspect of the system's evolution, the frequently
debated `choice' between It\^{o} or Stratonovich integration,
would be irrelevant because the corresponding Langevin equation
for either approach would be defined to ensure consistency with
the underlying master equation.

Hopefully the considerations above will motivate dedicated
experiments and direct simulations of turbulence to underpin a
rigorous formulation of the required stochastic equations, which
are likely to play an important role in future applications.

\section{Acknowledgements}

This work was supported by the Engineering and
Physical Sciences Research Council [grant number EP/V033883/1]
as part of the [D$^{*}$]stratify project.

\bibliographystyle{rs}
\bibliography{ref}

\end{document}